\documentclass[12pt,a4paper]{article}%%%%where IET is the template name
\usepackage[a4paper,margin=1in]{geometry}
\newtheorem{theorem}{Theorem}{}
{}
\newtheorem{remark}{Remark}{}
\usepackage[dvips]{graphicx}
\usepackage{color}
\usepackage{cite}
\usepackage{hyperref}
\usepackage[cmex10]{amsmath}
\usepackage{algorithmic}
\usepackage{url}
\usepackage{algorithm}
\usepackage{subfigure}
\usepackage{amsfonts}

\begin{document}
%
%\supertitle{IET Generation, Transmission and Distribution}
%
%
%\title{Inter-Area Oscillation Damping With~Non-Synchronized Wide-Area Power System Stabilizer}
%%\maketitle%
%   
%\author{\au{Abhilash Patel$^{1}$}, \au{Sandip Ghosh$^{2\corr}$}, \au{Komla A. Folly$^{3}$}}
%
%\address{\add{1}{Department of Electrical Engineering, Indian Institute of Technology Delhi, India}
%\add{2}{Department of Electrical Engineering, Indian Institute of Technology (BHU), Varanasi, India}
%\add{3}{Department of Electrical Engineering, University of Cape Town, Cape Town, South Africa}
%\email{sghosh@eee.iitbhu.ac.in}}
\title{Inter-Area Oscillation Damping With Non-Synchronized Wide-Area Power System Stabilizer}
\author{Abhilash Patel$^{1}$, Sandip Ghosh$^2$, Komla A. Folly$^3$\\
{\small $^1$Department of Electrical Engineering, Indian Institute of Technology Delhi, India}\\
{\small $^2$Department of Electrical Engineering, Indian Institute of Technology (BHU), Varanasi, India}\\
{\small $^3$Department of Electrical Engineering, University of Cape Town, Cape Town, South Africa}\\
{\small Emails: $^1$abhilash.patel@ee.iitd.ac.in, $^2$sghosh@eee.iitbhu.ac.in
, $^3$komla.folly@uct.ac.za }}
\date{}
\textit{{\color{blue}Accepted to IET Generation, Transmission and Distribution }}{\color{blue}}

doi:  10.1049/iet-gtd.2017.0017\\
{\let\newpage\relax\maketitle}
\begin{abstract}
One of the major issues in an interconnected power system is the low damping of inter-area oscillations which significantly reduces the power transfer capability. A speed deviation based Wide-Area Power System Stabilizer (WAPSS) is known to be effective in damping inter-area modes which uses feedback from remote locations. However, involvement of wide-area signals gives rise to the problem of time-delay, which may degrade the system performance. The delay in synchronized and non-synchronized feedback to WAPSS can have alternate performances. The effect of delays in such WAPSS with two types of feedback are studied and controllers are synthesized using $H_\infty$  control with regional pole placement to ensure adequate dynamic performance. To show the effectiveness of the proposed approach, two power system models have been used for the simulations. It is shown that the controllers designed based on the non-synchronized signals is more robust to time delay variations than the controllers  using synchronized signal.
\end{abstract}
\maketitle

\section{Introduction}

Modern interconnected power systems are very large and span over wide geographical area. Consequently, they inherit complex nonlinear dynamics, which are poorly damped. As a result, any changes in operating conditions may lead to electromechanical oscillations. Generally power system oscillations have multiple modes owing to the large-scale and interconnected dynamics. Such oscillations affect power system stability and performances. Usually, the inherent damping of the system  is not adequate to mitigate the oscillations. 

The oscillations are characterized through the swing modes of the system and these can be categorized as \textit{local} and \textit{inter-area modes}~\cite{kundur94book}. For local modes, either an individual or a group of machines within an area oscillates against another individual or a group of machines from  the same area. On the other hand, for inter-area modes, a group of machines from an area oscillates against group of machines from another area, usually connected over a weak tie-line. The local modes have stronger impact on the states of local-area machines and, hence, these can be effectively damped using local signals and controllers, such as using local Power System Stabilizers (PSSs). On the contrary, the inter-area modes involving multiple areas are difficult to control using only local signals. It may also happen that the inter-area modes are better controllable from one area but better observable in different area signals~\cite{chow00}. This rises the issue of wide-area control of the inter-area modes using signals from different areas.

The limitation on availability of feedback signals from different areas is somewhat relaxed with the deployment of wide-area measurement systems over wide geographical areas, and remote signals are now available for feedback in order to realize Wide-Area Power System Stabilizer (WAPSS). Using both the local and remote signals as feedback to the controller, the inter-area modes can be better damped. It has also been reported that WAPSS takes 4 to 20 times lesser control effort to adequately damp inter-area modes compared to local PSS~\cite{kamwa05}. 

{Several studies in the literature have explored the benefits of using WAPSSs. In fact, several design methods have been proposed for designing WAPSSs. A phase compensating lead-lag compensator has been designed as WAPSSs~\cite{zhang13,aboul96}. Such a design is simple and easy to implement, but it performs non-uniformly under varying operating conditions. With the help of modern control theory, adaptive and robust controllers have been designed to address the above robustness problem ~\cite{padhy12,hui02,kamwa01,chaudhuri04b,zolotas07,xie06,zhang08,aranya11,majumder05}. Adaptive-WAPSS uses online estimation of system parameters and computation of control law adaptively as in~\cite{xie06,aranya11,chaudhuri04b}. Robust control methods have unique superiority in dealing with the uncertainty in system parameters and external perturbations. To cope with bounded operating conditions, robust control theory comes as an effective tool for controller synthesis. A linear quadratic gaussian controller has been used in~\cite{zolotas07,bhadu2016robust} along with loop transfer recovery scheme to achieve guaranteed robustness. $H_\infty$ loop shaping approach along with regional pole placement has been proposed for wide-area controller design in~\cite{majumder05}. In \cite{zhang08,chen06}, the design of $H_2/H_\infty$ mixed sensitivity controller with regional pole placement criterion has been designed.  Here, the controller is synthesized to minimize weighted sensitivity with critical modes of the closed-loop system placed in a pre-specified region in the left-half complex plane ensuring certain dynamic performance.}

%Several works address the time delays in wide area control loop explicitly\cite{yao09,yao11,yao14,yao15,mokhtari13,hashmani11,chaudhuri04,majumder05b}.A model-free fuzzy controller, based upon operator experience rule base, has been designed in \cite{mokhtari13}. In this, the output membership functions are shifted in a manner to compensate the effect of the delay. A lead-lag compensator to compensate the phase is proposed in \cite{yao11,yao14} where the gain of the compensator is calculated from a tradeoff between delay margin and damping. To handle the uncertainty in delay \cite{hashmani11} proposed a linear fractional transform model to be used in $H_\infty$ control framework. Predictor based approach also used in \cite{yao09,yao15,chaudhuri04,majumder05b},where\cite{chaudhuri04,majumder05b} uses unified smith predictor to overcome minimum damping issue with classical smith predictor and \cite{yao09,yao15} uses generalised predictive control along with RLS base model identification to make system adaptive in nature.}
%The idea has been implemented in literature with robust control theory~\cite{zhang08,majumder05}, adaptive control algorithm~\cite{aranya11,zhao16}, evolutionary and learning methods~\cite{wang14,shakarami2016wide,hadidi2013reinforcement} advocated the proficiency.

Though the use of wide-area signals provides better dynamic behavior, the involvement of remote signal introduces challenges such as time-delay in the process of acquiring the remote signals. Such time-delay in the system may reduce the damping or even lead to instability~\cite{wuhe04,ghosh16}. The time-delay can be compensated using predictor approach~\cite{yao15,chaudhuri04}, adding phase lead~\cite{zhang2012adaptive} or developing a stochastic time-delay expectation model approach~\cite{zhang16new}. The control gain can also be tuned by exploiting the trade-off between delay margin and damping \cite{yao14}. However, the design methods lack the robustness to time-varying delay or are computationally expensive.

%{\color{red}
%We have already established that there exists feedback configuration which can enhance the delay tolerability \cite{ghosh16}. The work here presented as of tutorial fashion covering all the stpdf in the design. At first, $H_\infty$ with regional pole placement controllers are designed without considering delay in design. Then, affect of delays in feedback are studied in nonlinear simulation. The non-synchronized signals which shows better tolerability in the analysis, is used again to re-design controllers. The delay is approximated with Pade's model, and shown that controller designed with non-synchronized feedback is robust to delay. The concept is validated in two case studies, two-area system and new England system.
%}
{
One important issue while employing speed-based WAPSSs that use speed difference between machines from different areas as feedback signals is whether to use synchronized signals or not. It has been shown that there exists a feedback configuration which can enhance the delay tolerability by better utilizing the local signal (non-synchronized signal) than using the synchronized one~\cite{ghosh16}. In this work, the robustness of delay has been analyzed based on the small signal model through computation of eigenvalues for delay systems. However, the work in \cite{ghosh16} does not deal with WAPSSs synthesis. Note that, often, the advanced controller design, such as $H_\infty$ controller, may not yield the same performance as it is observed through analysis due to their inherent conservativeness (lack of existance of necessary and sufficient conditions). Therefore, comparing the designed controller performances for synchronized and non-synchronized use of feedback signals in speed-based WAPSSs is important.

In this paper, design of speed-based WAPSSs is carried out employing the second-order Pade approximation of the delay. The WAPSS is designed for the following cases: (i) Not considering delay in design and (ii) Considering delay in design. $H_\infty$ control design method with regional pole placement is employed for the controller design. The effect of time-delay for synchronized and non-synchronized feedback using nonlinear simulations is demonstrated. It is shown through comparison that the design yields also corroborate the analysis carried out in \cite{ghosh16}. Two case studies, the two-area system and the new England system, are considered for demonstrating the effectiveness of the designed WAPSSs.}

The paper is organized as follows. The next section introduces the control problem formulation for both synchronized and non-synchronized configuration. Section III presents the simulation results where WAPSSs are designed for the two-area 4 machines IEEE benchmark model. In section IV, the results for a ten-machine New England system, a larger system with multiple inter-area modes, are tested. Section V presents the conclusion of this paper.

\section{Control Problem Formulation}
This section presents the synchronized and the non-synchronized feedback configurations for speed-based WAPSS. The corresponding system models are presented that are used for WAPSS design.
\subsection{System with Synchronized and Non-Synchronized Feedback}
\begin{figure}[!tbh]
\centering
\includegraphics[width=3in]{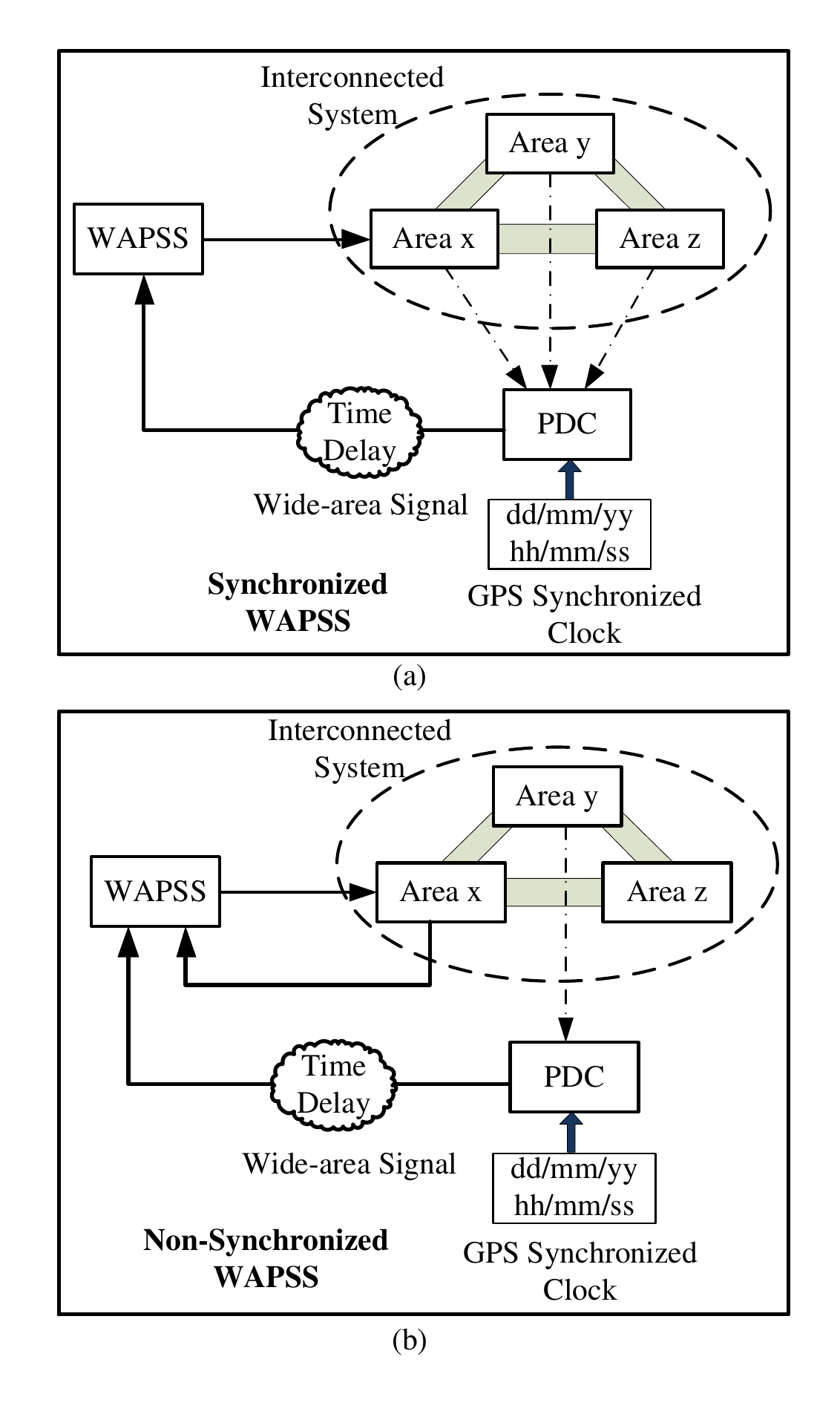}
 \caption{a) Configuration for synchronized WAPSS. b) Configuration for non-synchronized WAPSS.}
\label{fig:WAPSS_structure}
\end{figure}

System with two-loop wide-area control structure using PSS is shown in Fig. \ref{fig:WAPSS_structure}. It may be noted that the inter-area oscillation is well observed through the speed difference of the $i^{th}$ and $j^{th}$ generators from the two different areas (i.e.
$\Delta\omega_{ij}=\Delta\omega_i-\Delta\omega_j$, $\Delta\omega$ representing the speed deviation) \cite{aboul96}. However, in order to facilitate the synchronization for better observation, an equal amount of delays are introduced in the synchronized signals as in Fig. \ref{fig:WAPSS_structure} (a), whereas if the two signals are used without synchronization as in Fig. \ref{fig:WAPSS_structure} (b) then the delay in the local signal $\Delta\omega_i$ is negligible. 
%\begin{figure}
%\centering
%\includegraphics[width=3in]{Figures/WPSS_structure.eps}
% \caption{WAPSS configuration at $i^{th}$ generator terminal}
%\label{fig:WAPSS_structure}
%\end{figure}

To incorporate the above two configurations in system model, consider the linearized power system model:
\begin{equation}
\label{eq:plant}
\begin{split}
\dot{x}_p&=A_px_p+B_ww+B_pu_p,\\y_p&=C_px_p,\\z&=C_zx_p+D_zw,
\end{split}
\end{equation}
where $x_p$ is the state, $u_p$ is the wide-area control signal fed to the exciter, $y_p$ is the measured output considered here as wide-area signals, $z$ is the desired output for which the controller design performance is chosen, $A_p,\ B_w,\ B_p,\ C_z,\ D_z,\ C_p$ are time-invariant matrices of appropriate dimension.
The WAPSS controller dynamics is considered as: 
\begin{equation}
\label{eq:controller}
\begin{split}
\dot{x}_c&=A_cx_c+B_cu_c,\\y_c&=C_cx_c,
\end{split}
\end{equation}
where $x_c$ is the state of the controller, $u_c$ is the wide-area feedback to the controller, delayed or non-delayed ones as the case will be, $y_c$ is the control signal fed to the system (\ref{eq:plant}), i.e. $u_p$, $A_c,\ B_c,\ C_c$ are the controller matrices to be designed. The plant and controller transfer functions can be written as $G_p(s)=C_p(s)(sI-A_p)^{-1}B_p$ and $G_c(s)=C_c(s)(sI-A_c)^{-1}B_c$, respectively. 

For controller design, time-delay present in the remote signal is modeled with second-order Pade approximation. A discussion on the second-order modeling of the delay is presented in the next section. State-space model of Pade-approximated delay term is as:
\begin{equation}
\label{eq:pade2nd}
\begin{split}
\dot{x}_d&=A_dx_d+B_du_d,\\y_d&=C_dx_d+D_dy_d,
\end{split}
\end{equation}
where $u_d$ is the wide-area signal without delay, $x_d$ is states of the approximated delay model, $y_d$ is the delayed wide-area feedback to the controller, and constant matrices $A_d$, $B_d$, $C_d$ and $D_d$ are given as follows
\begin{equation*}
A_d=\begin{bmatrix}
0& 1\\
\frac{-12}{{T_d}^2}&\frac{6}{T_d}\\
\end{bmatrix},
B_d=\begin{bmatrix}
0\\
1\\
\end{bmatrix},
C_d=\begin{bmatrix}
0 &-T_d\\
\end{bmatrix},
D_d=\begin{bmatrix}
1\\
\end{bmatrix}.
\end{equation*}
where $T_d$ is the time-delay.

Next, (\ref{eq:plant})-(\ref{eq:pade2nd}) are used to represent the closed-loop system structures corresponding to the synchronized and the non-synchronized configurations of Fig. \ref{fig:WAPSS_structure}.
\subsubsection*{Synchronized Feedback:} \quad 
For synchronized feedback in Fig. \ref{fig:WAPSS_structure}(a), both the local and the remote signals are time-synchronized and, therefore, an equal amount of delays are introduced in both the signals. The state space representation of the closed-loop system is as:
\begin{equation*}
\label{eq:closed_loop_syn}
\begin{split}
\dot{x}&=A_sx+B_ww,\\z&=C_zx_p+D_zw,
\end{split}
\end{equation*}
where $x={\begin{bmatrix}{x_p}^T&{x_d}^T&{x_c}^T
\end{bmatrix}}^T$ and  
\begin{equation*}
A_s=\begin{bmatrix}
A_p& 0&B_pC_c\\
B_dC_p&A_d&0\\
B_cD_dC_p&B_cC_d&A_c
\end{bmatrix}.
\end{equation*}
\subsubsection*{Non-synchronized Feedback:}\quad
For non-synchronized feedback in Fig. \ref{fig:WAPSS_structure}(b), the local signal is used as it is without time-delay. So the delay is observed only in the remote signal. Consider $C_p={\begin{bmatrix}
{C_l}^T &{C_r}^T \end{bmatrix}}^T$, $C_l$ and $C_r$ corresponds to local and remote signals respectively. Similarly, the state-space representation of the closed-loop system can be written as:
\begin{equation*}
\label{eq:closed_loop_syn}
\begin{split}
\dot{x}&=A_{ns}x+B_ww,\\z&=C_zx_p+D_zw,
\end{split}
\end{equation*}
where $x={\begin{bmatrix}{x_p}^T&{x_d}^T&{x_c}^T
\end{bmatrix}}^T$ and 
\begin{equation*}
A_{ns}=\begin{bmatrix}
A_p&0&B_pC_c\\
B_dC_r&A_d&0\\
B_cC_l+B_cD_dC_r&B_cC_d&A_c
\end{bmatrix}.
\end{equation*}
As the inter-area oscillations are well observable in the difference if speed deviation of two machines from the different areas, one can opt for the performance measure, $z$, to be as $\Delta\omega_{ij}$ with suitable weight augmentation.

\subsection{Controller Design}
The $H_\infty$ controller design with regional pole placement is used in this work for comparison of controller performances for non-synchronized and synchronized signals. The controller design method is presented next.
\subsubsection{$H_\infty$ Control}
\label{H_inf_Control_section}
\begin{figure}[!tbh]
\centering
\includegraphics[width=3in,trim=2in 1.5in 1.5in 2in,clip=true]{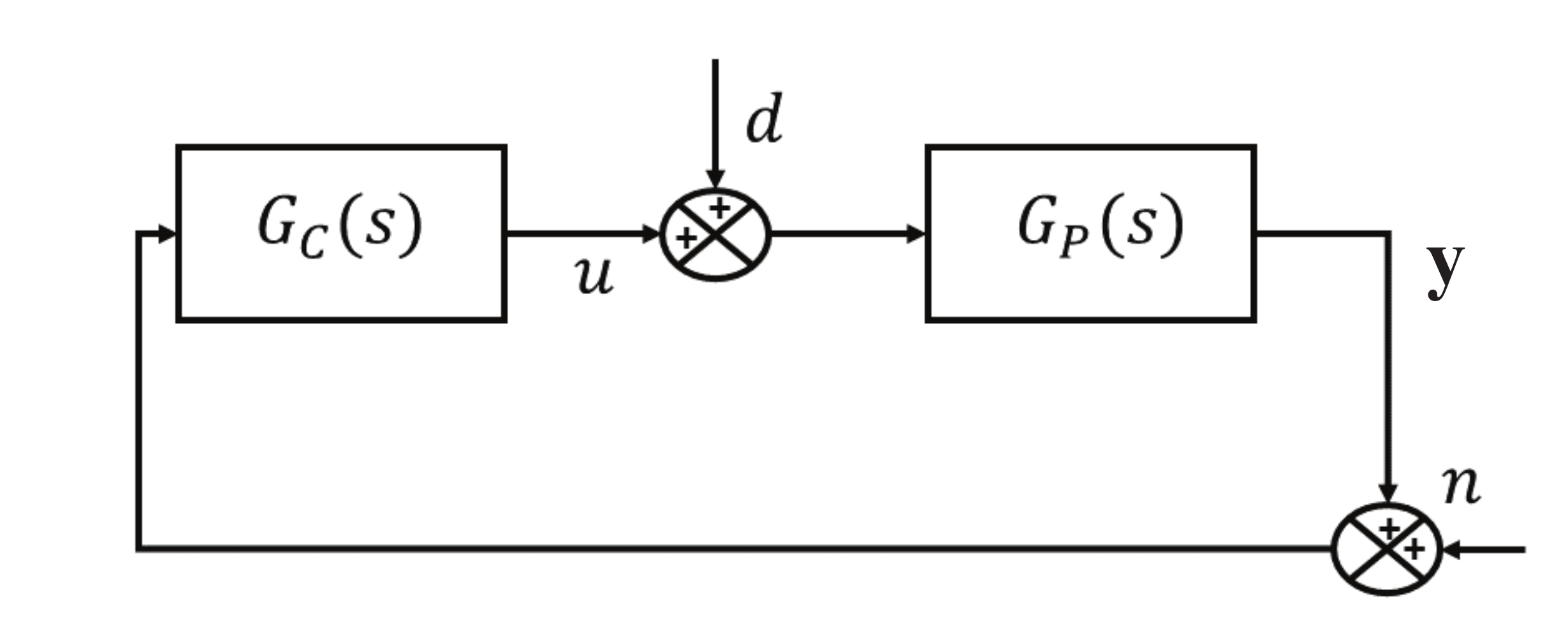}
 \caption{Feedback control structure for controller design}
\label{fig:control_blk_diag1}
\end{figure}

A block diagram representation of robust control problem is presented in Fig. \ref{fig:control_blk_diag1}. Here, $G_p$ is the system as in \ref{eq:plant} is to be controlled with the $G_c$ as in \ref{eq:controller} in the presence of disturbance ($d$) and noise ($n$).

From Fig.\ref{fig:control_blk_diag1}, the output $y$ can be written as:
\begin{equation}\label{eq1}
y=(1-G_pG_c)^{-1}G_pd+G_pG_c(1-G_pG_c)^{-1}n ,
\end{equation} 
where $(1-G_pG_c)^{-1}$ is called the sensitivity function $S$ and $G_pG_c(1-G_pG_c)^{-1}$ is the complementary sensitivity function $T$ \cite{skogestad14}. To have a good disturbance rejection $S$ should be small over a frequency range where disturbance rejection is required, whereas $T$ should be small to attenuate noise at the frequency range where noise attenuation is required. However, $T$ and $S$ are complimentary to each other since $S+T=1$. It is well known that disturbances are of low frequency in nature whereas noises are of high frequency. With the help of frequency-dependent weights, one can set an objective exploring their trade-off. 

Considering $H_\infty$ gain as the performance measure ($\|\cdot\|_\infty$ is the $H_\infty$ norm of the
transfer function, i.e. the supremum of its largest singular value over
frequency, denoted as $\gamma_\infty$), one can define the performance requirement as~\cite{skogestad14}:
\begin{equation}
\left\|\begin{array}{c}W_1S\\W_2T\end{array}\right\|_\infty\le\gamma_\infty,
\end{equation}
where $W_1$ and $W_2$ are suitable weight functions that provide the flexibility of exploring the frequency based design trade-off
between $S$ and $T$. Usually $W_1$ is chosen as a low pass filter whereas $W_2$ is chosen as a high pass filter.
%%%%%%%%%%%%%%%%%%%%%%%%%%%%%%%%%%%%%%%%%%
%Now, $A$ is stable and $H_\infty$ norm of (\ref{eq:Hinf_Closedloop_System}) is smaller than $\gamma$ if and only if there exists a symmetric $P$ satisfying
%\begin{equation}\label{Hinf_LMI} 
%  \left( {\begin{array}{ccc}
%   {A}^TP+PA & PB&C^T \\
%    {B}^TP & -\gamma I & D \\
%    C &  D&-\gamma I  \\
%  
%  \end{array} } \right)
%  < 0.
%\end{equation}
%%%%%%%%%%%%%%%%
%Representation of the disturbance rejection problem as uncertain system problem as a robustness one with additive uncertainty is shown in Fig.\ref{fig_blk_diag2}, where $\Delta(s)$ is the additive uncertainty. Weight function can be adjusted as per small gain theorem to achieve robustness~\cite{gahinet94}. Although mixed sensitivity control framework can handle disturbance rejection and robustness very well, it lacks the control over the transient behavior of the system~\cite{skogestad14}.
%\begin{figure}[!tbh]
%\centering
%\includegraphics[width=3in]{Figures/blk_diag2.pdf}
% \caption{Generalised control scheme for system with uncertainty}
%\label{fig_blk_diag2}
%\end{figure}
\subsubsection{Regional Pole Placement}
While $H_\infty$ control addresses the frequency domain specifications in terms of disturbance rejection and robustness effectively but it lacks control over the transient performance. A desired response can be achieved by forcing the closed-loop poles to be placed in a pre-specified Linear Matrix Inequality (LMI) region. Moreover, since damping improvement is the main concern here, one requires to ensure, the minimum desired damping at the design stage.

One can describe a region in complex plane with LMI equations~\cite{chilali96}. A conical sector is constructed as per minimum damping requirement ($\cos 2\theta=\zeta$, $\zeta$ is the damping factor and $\theta$ is the ineer angle in the complex plane). The characteristic function of conic sector with apex at origin and inner angle of $2\theta$ can be expressed as
\[
   f_D(v)=
  \left[ {\begin{array}{cc}
   \sin\theta(v+\bar{v}) & \cos\theta(v-\bar{v}) \\
   \cos\theta(\bar{v}-v) & \sin \theta(v+\bar{v}) \\
  \end{array} } \right],
\]
where $v$ represents pole location and $\bar{v}$ is its complex conjugate. By exploiting this characteristic function, regional pole placement along with $H_\infty$ control can be carried out in multi-objective output feedback framework \cite{scherer97} as given in Theorem 1 below.
\begin{theorem}
\label{Theorem}
Given a constant $\gamma >\ 0$ and a linear time-invariant plant (\ref{eq:plant}), there exists an output feedback controller (\ref{eq:controller}), such that $\|T_{zw}|_\infty\le\gamma$ and all its closed-loop eigenvalues resides in the region in left half plane with conical sector with inner angle $2\theta$ and apex at origin, if there exist symmetric matrices $R$ and $S$, matrices $\hat{A}$, $\hat{B}$, $\hat{C}$, satisfying the below LMI constraints:
\begin{equation}
\begin{bmatrix}
\Phi_1&*&*&*\\
\hat{A}+A'&\Phi_2&*&*\\
B_w'&RB_w'&-\gamma I&*\\
C_zS&C_z&D_z&-\gamma I
\end{bmatrix}<0
\end{equation}
\begin{equation}
\begin{bmatrix}
\sin\theta(\Phi_3+\Phi_3 ')&\cos\theta(\Phi_3-\Phi_3 ')\\
\cos\theta(\Phi_3'-\Phi_3)&\sin\theta(\Phi_3 '+\Phi_3)
\end{bmatrix}
<0
\end{equation}
\begin{equation}
\begin{bmatrix}
S&I\\
I&R
\end{bmatrix}
>0
\end{equation}
where
\begin{equation*}
\begin{split}
\Phi_1 &= AS+SA'+B\hat{C}+(B\hat{C})'\\
\Phi_2 &= A'R+RA+\hat{B}C+(\hat{BC})'
\end{split}
\end{equation*}
\begin{equation*}
\Phi_3 =
\begin{bmatrix}
AS+B\hat{C}&A\\
\hat{A}& RA+\hat{B}C
\end{bmatrix}
\end{equation*}
\end{theorem}
As the matrices $\hat{A},\ \hat{B},\ \hat{C},\ R,\ S$ are present affinely in the above inequalities, it can be solved with a LMI solver. Once these matrices are obtained, the controller (\ref{eq:controller}) can be constructed using the following:
\begin{equation*}\begin{split}&\hat{A}=SA_pR+SB_pC_cU^T+VB_cC_pR+VA_cU^T,\\ &\hat{B}=VB_c, \quad\hat{C}=C_cU^T\end{split}\end{equation*} for matrices $U$ and $V$ satisfying $UV^T=I-RS$.

Based on the synchronized or non-synchronized feedback considered, the plant (\ref{eq:plant}) and the corresponding controller (\ref{eq:controller}) can be considered and designed. It may be noted that the delay model (\ref{eq:pade2nd}) is augmented to the plant dynamics appropriately based on whether it is considered in the design or not.
{
\begin{remark}
There are possibilities that the LMI condition in Theorem 1 may yield infeasibile solution if the design is subjected to hard design constraints. Approaches to handle the situation are: (i) to relax the imposed design constraints (for example, reducing the desired damping value),  (ii) to find achievable performance posing the problem as an optimization problem (maximizing damping factor). However, for the case studies carried out in this work, we could obtain feasible solutions for several different damping factor values.
\end{remark}
}

Next, we present two case studies comparing the designs for synchronized and non-synchronized WAPSSs with particular emaphasis on the effect of delay variations.
\section{Case Study-I}
\label{Case_Study1}
\begin{figure}[!tbh]
\centering
\includegraphics[width=3.5in]{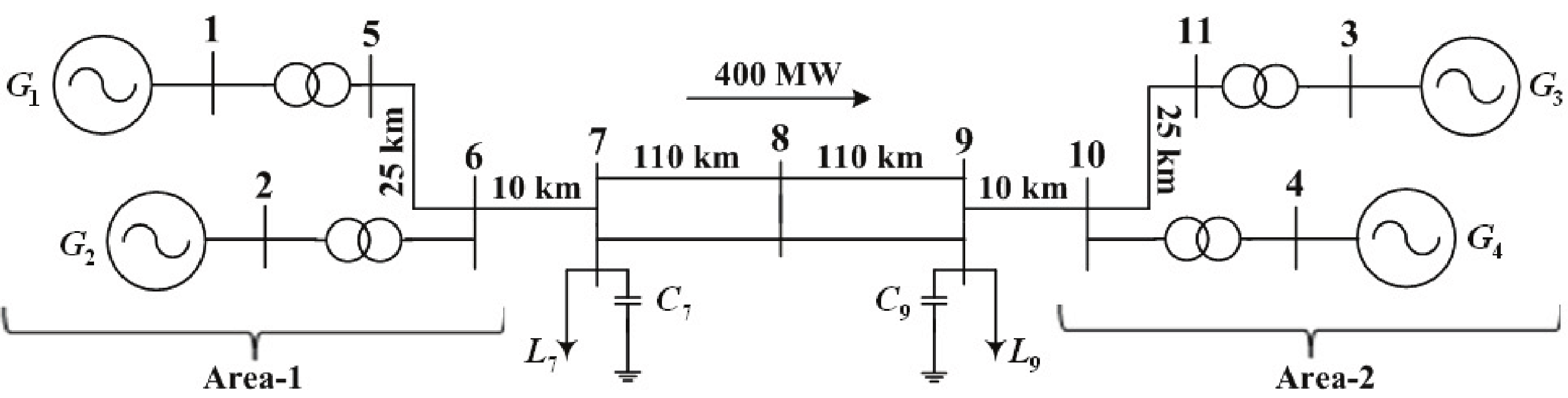}
 \caption{The 4 Machine 11 Bus study system~\cite{kundur94book}}
\label{fig:4M11B}
\end{figure}
\begin{figure*}[b]
\centering
 \subfigure[]
    {%
     \includegraphics[width=3in,height=1.8in]{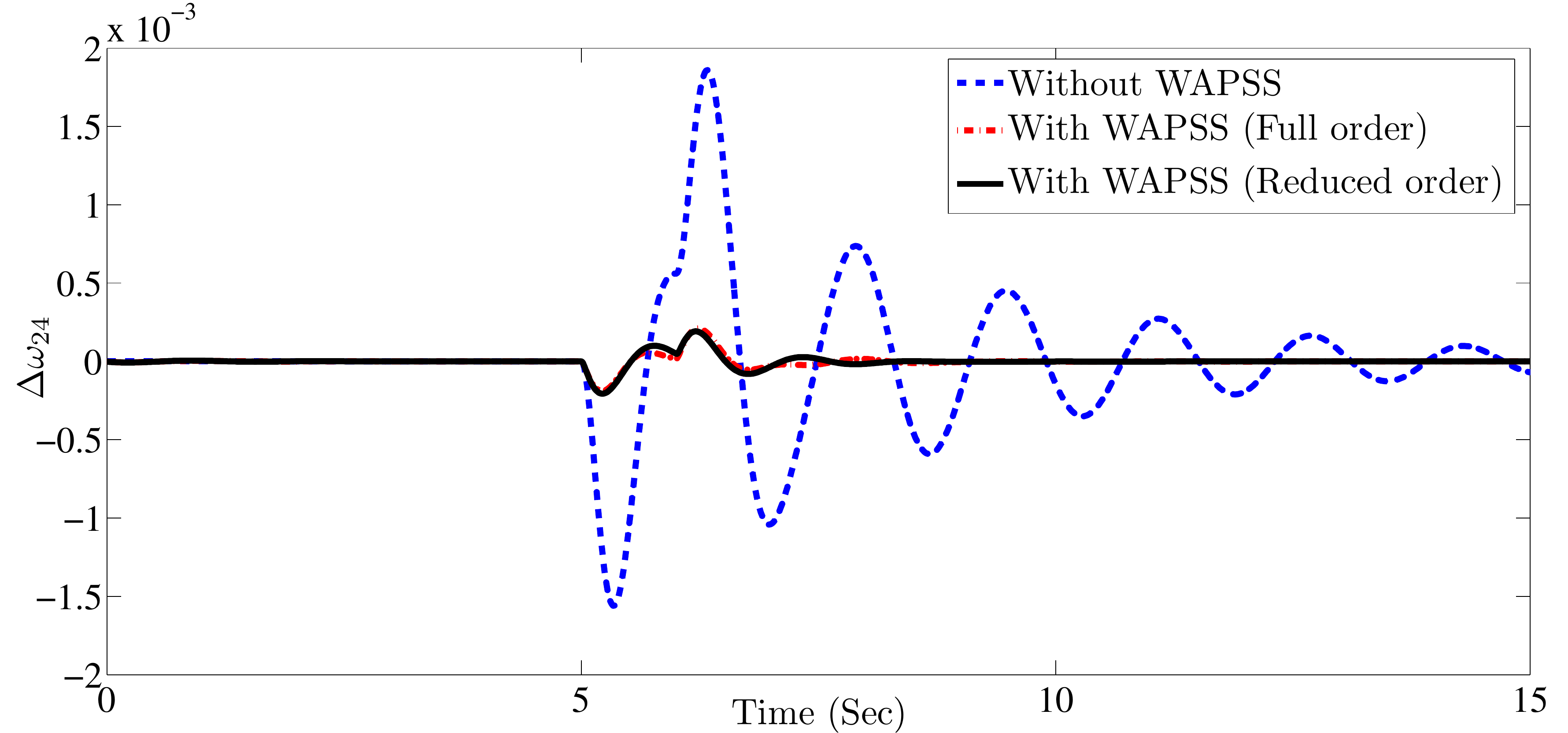}
     \label{fig:dW24_WAPSS}
    }
    \subfigure[]
    {%
    \includegraphics[width=3in,height=1.8in]{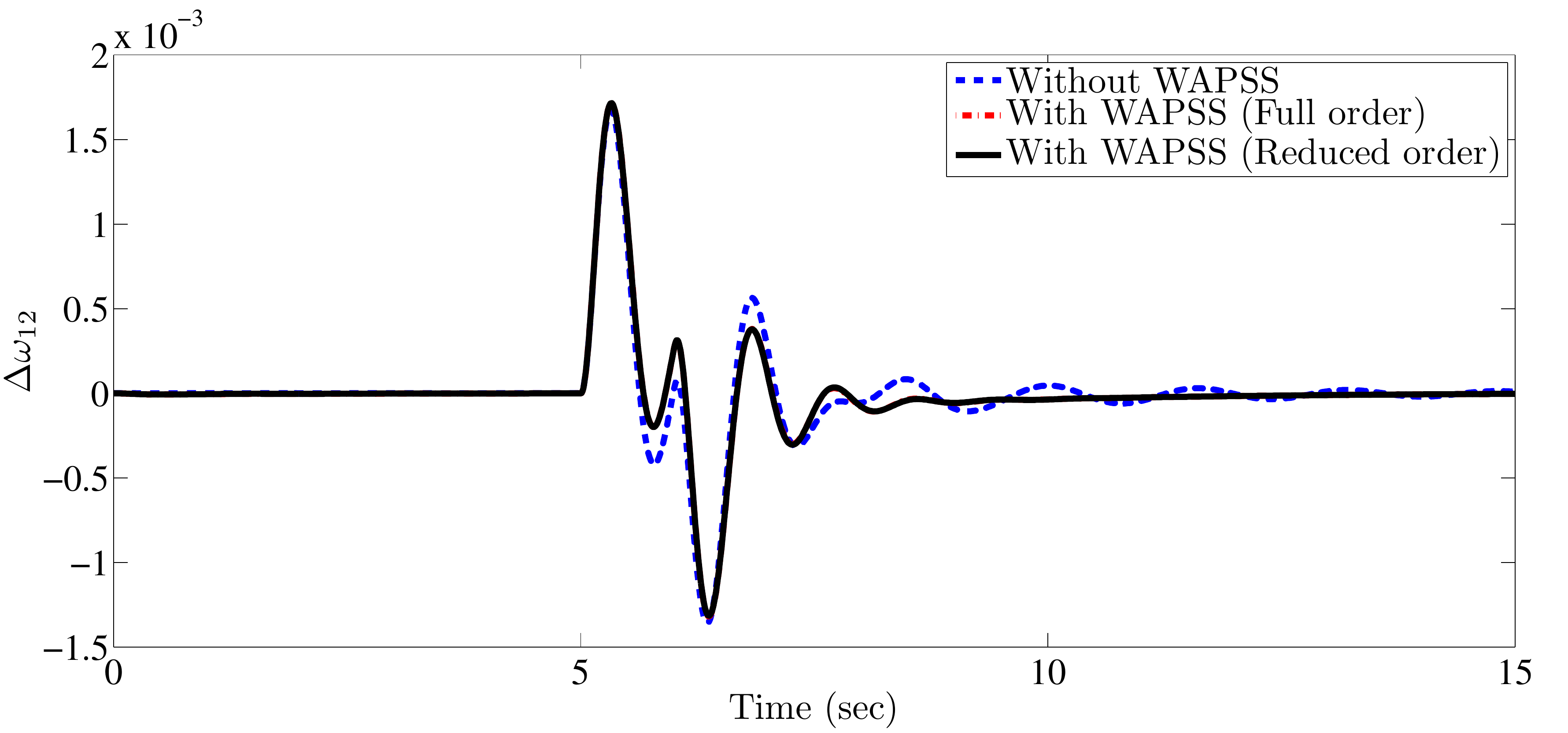}
    \label{fig:dW12_WAPSS}
    }
    \subfigure[]
    {%
    \includegraphics[width=3in,height=1.8in]{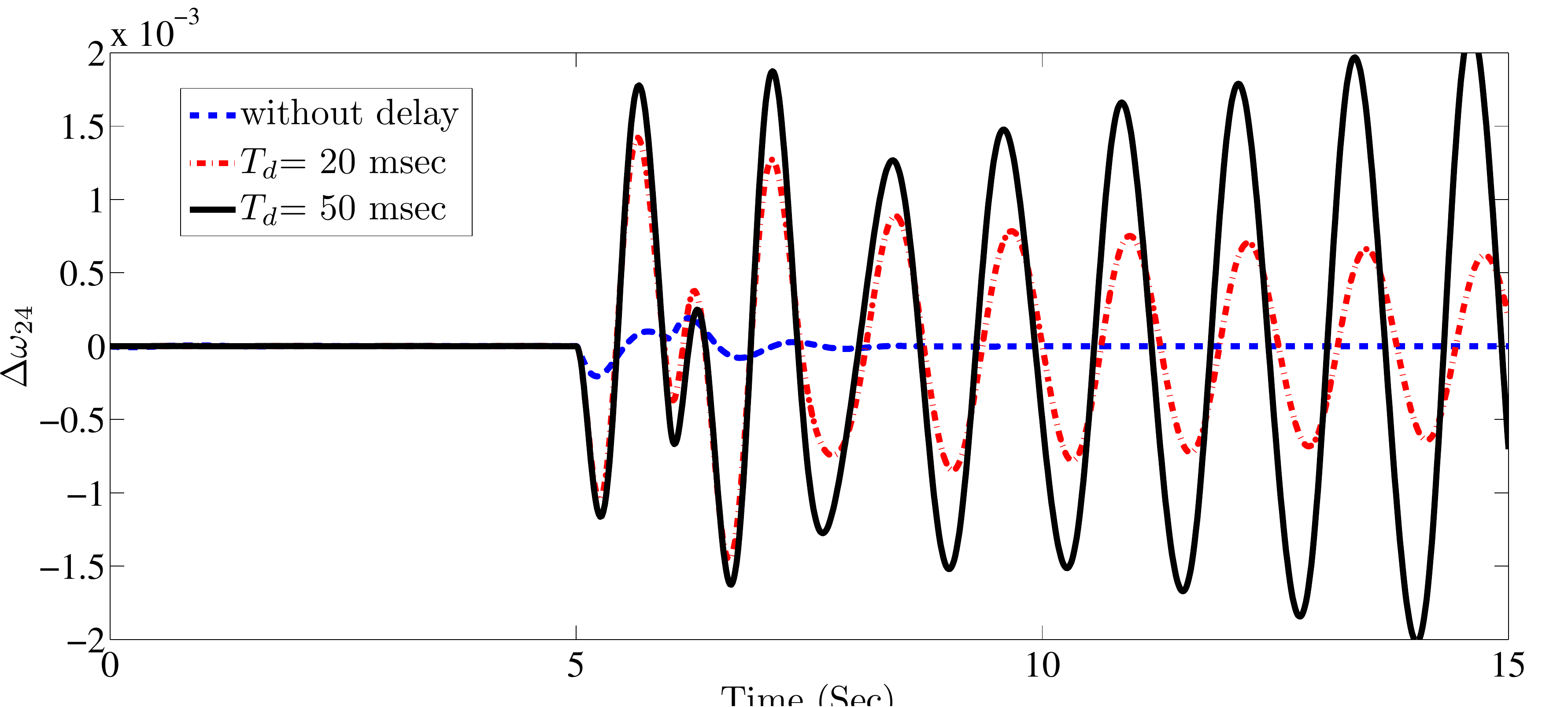}
\label{fig:dW24_alldel_syn}
    }
    \subfigure[]
    {%
     \includegraphics[width=3in,height=1.8in]{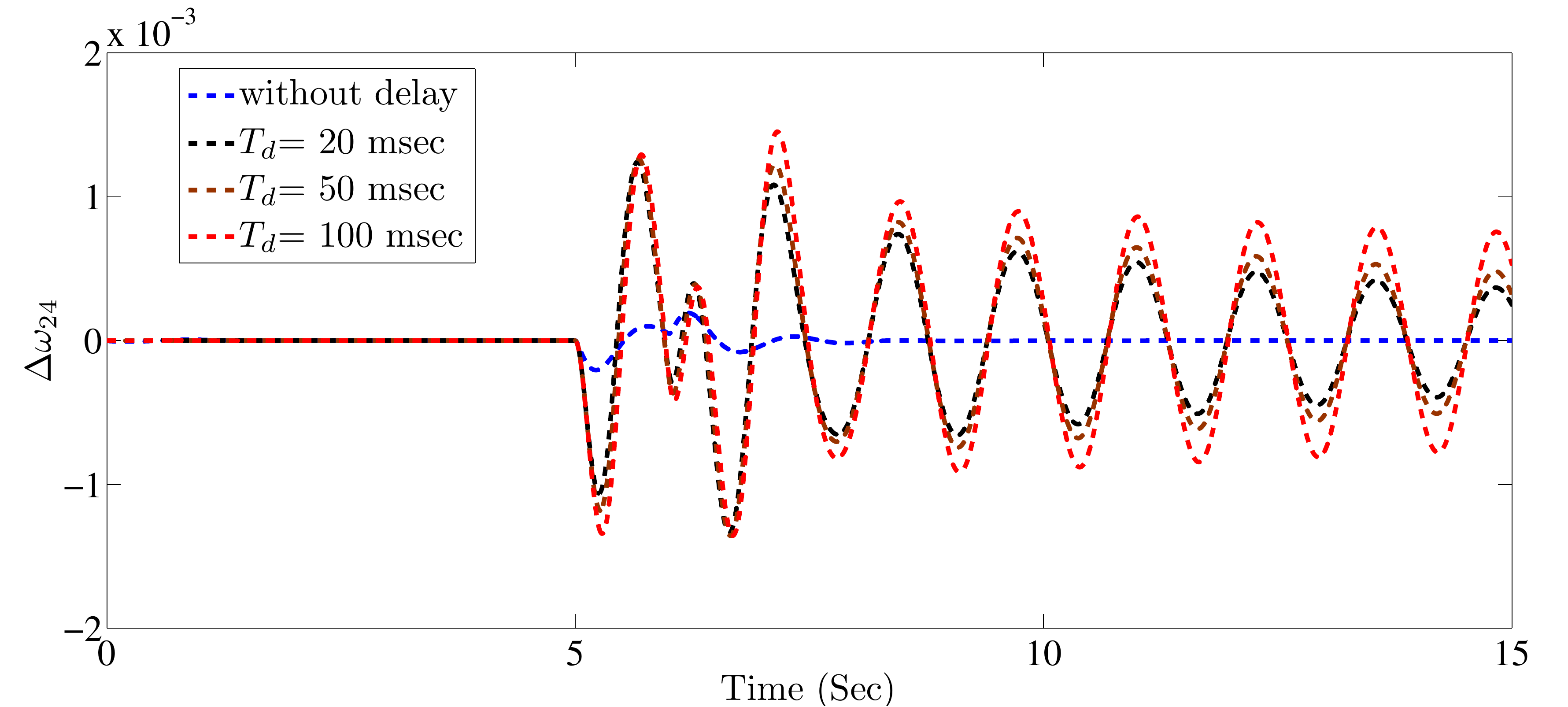}
    \label{fig:dW24_alldel_nonsyn}
    }
    \caption{a) $\Delta\omega_{24}$ variation: \color{blue}- - without WAPSS, \color{red}$-\cdot-$with full order WAPSS,\color{black}--- with WAPSS reduced order.(b) $\Delta\omega_{12}$ variation: \color{blue}- - without WAPSS, \color{red}$-\cdot-$with full order WAPSS,\color{black}--- with WAPSS reduced order.(c) $\Delta\omega_{24}$ variation for different delay ($T_d$) in wide-area loop.(d) $\Delta\omega_{24}$ variation for different time-delay ($T_d$) in remote signal.}
%  \label{fig_dW34vsGain}
\end{figure*}

The IEEE benchmark two-area system developed for study of inter-area
oscillations~\cite{kundur94book} is first used to demonstrate the proposed WAPSS design procedure and its effectiveness.

The system consists of 11 buses and 4 generators where two areas are connected by a weak tie-line. Each area of the system contains two generators equipped with local PSSs at $G_1$ and $G_3$
to provide sufficient damping to the local modes. The nominal system parameters and operating point without any
wide-area control are considered as in ~\cite{kundur94book,Xtreport}.

The system is modeled in MATLAB-Simulink and linearized around an operating point of $400 MW$ tie-line transfer. The linearized model is obtained as $58^{th}$ order. The modes of the system are then studied with modal analysis. Three modes $M_1, M_2, M_3$ are identified as the swing modes with $M_1:-0.316\pm j3.91$ is the inter-area mode having poor damping. The wide-area loop is selected based on geometric measure for loop-selection corresponding to the inter-area mode. Through the geometrical measures as in ~\cite{hamdan88}, $\Delta\omega_{24}$ is chosen for feedback signal to the controller and $G_4$ is selected as the WAPSS location. As the  model is of large order, it is reduced to an $8^{th}$ order model. Note that, reduced order model retains the inter-area mode $M_1$ and a local mode $M_3$ of the original system, but loses $M_2$. 
\subsection{WAPSS Synthesis (Without Considering Delay)}
The controller is designed using the reduced order model without considering the delay in feedback. The controller is synthesized using procedure described in Section-II. The weight functions are selected as $W_1=\frac{10}{s+10}$, $W_2=\frac{100s}{s+10}$. To improve damping of the inter-area mode, a conical region is selected for pole placement with minimum damping ratio of $0.2$. The designed controller is of $10^{th}$ order due to extra states from weight functions, which is further reduced to a $5^{th}$ order using the Hankel norm reduction technique \cite{skogestad14}. From small signal analysis, the inter-area mode with WAPSS is found out to be $-1.02\pm j3.91$. Therefore, the damping is improved from $0.08$ to $0.253$.

Next, the WAPSS is validated through nonlinear simulation with a pulse change of $0.05\ pu$ in the voltage reference at $G_2$ terminal as a disturbance. The effectiveness of both the full-order and the reduced-order controllers are shown in Fig. \ref{fig:dW24_WAPSS} in terms of speed deviations. Without the wide-area loop, speed deviation is taking more than $10\ sec$ to get settled, which is undesirable following IEEE guidelines~\cite{kundur03}. With the designed wide-area controller, it can be seen that the inter-area oscillation damps out quickly as compared to the case without wide-area control. From Fig. \ref{fig:dW24_WAPSS}, it can be seen that the reduced-order controller works as good as the full-order controller. Therefore, the reduced order controller can be used which is relatively easier to implement. It is also observed that the controller in area-2 has no counter effect on $M_2$ which can be seen from variation in local mode of area-1 in Fig. \ref{fig:dW12_WAPSS}.

Next, to study the effect of delay, the system is simulated for different delays in the wide-area loop. A detailed linear analysis of the effect of delay considering both the synchronized and non-synchronized signal has been investigated in ~\cite{ghosh16}. First, considering  delays in the synchronized feedback, it can be seen from Fig. \ref{fig:dW24_alldel_syn} that damping reduces and the system becomes unstable  even for small delays.
\begin{figure*}[t]
\centering
    \subfigure[]
    {\includegraphics[width=3in,height=1.8in]{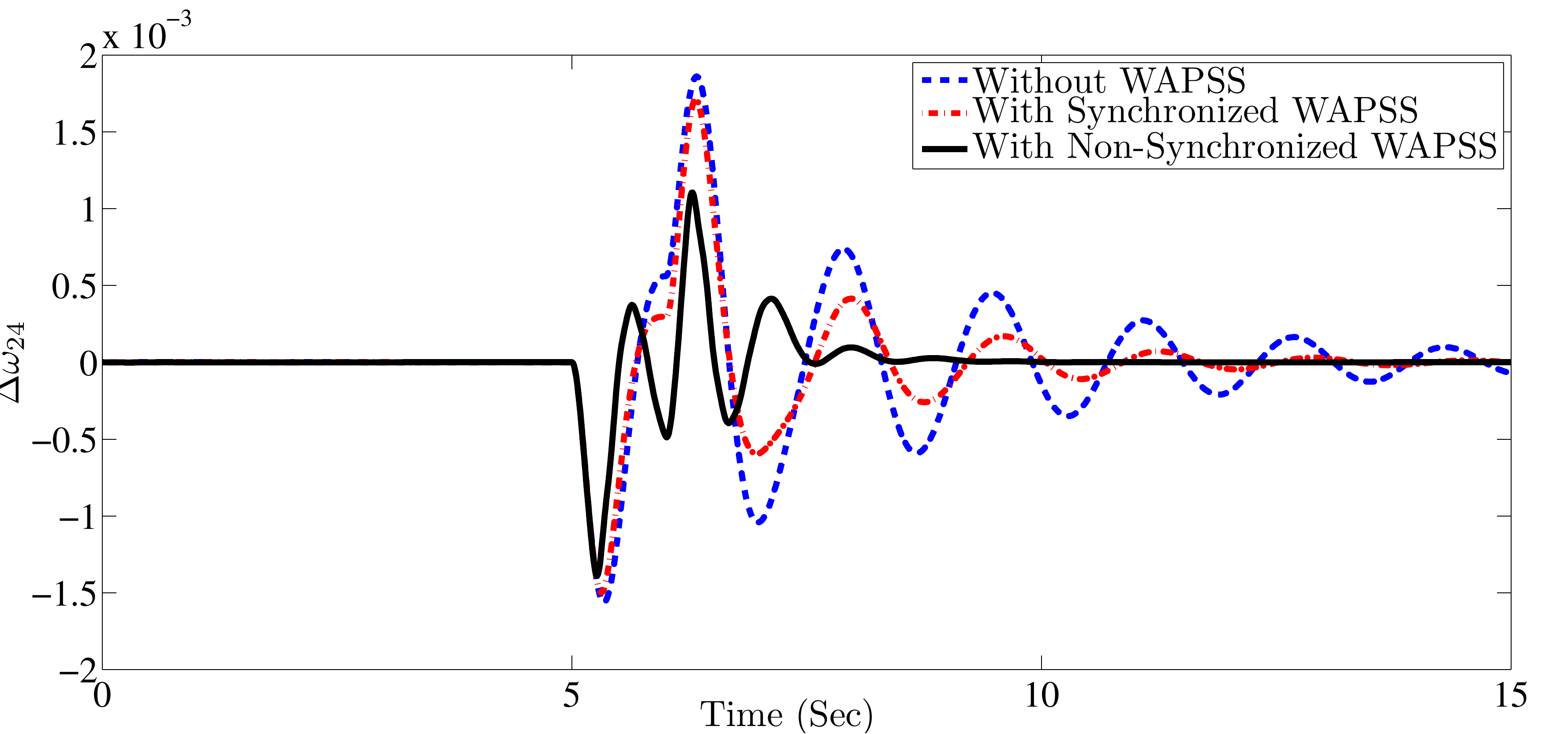}
    \label{fig:dW24_Syn_NonSyn}
    }
    \subfigure[]
    {%
    \includegraphics[width=3in,height=1.8in]{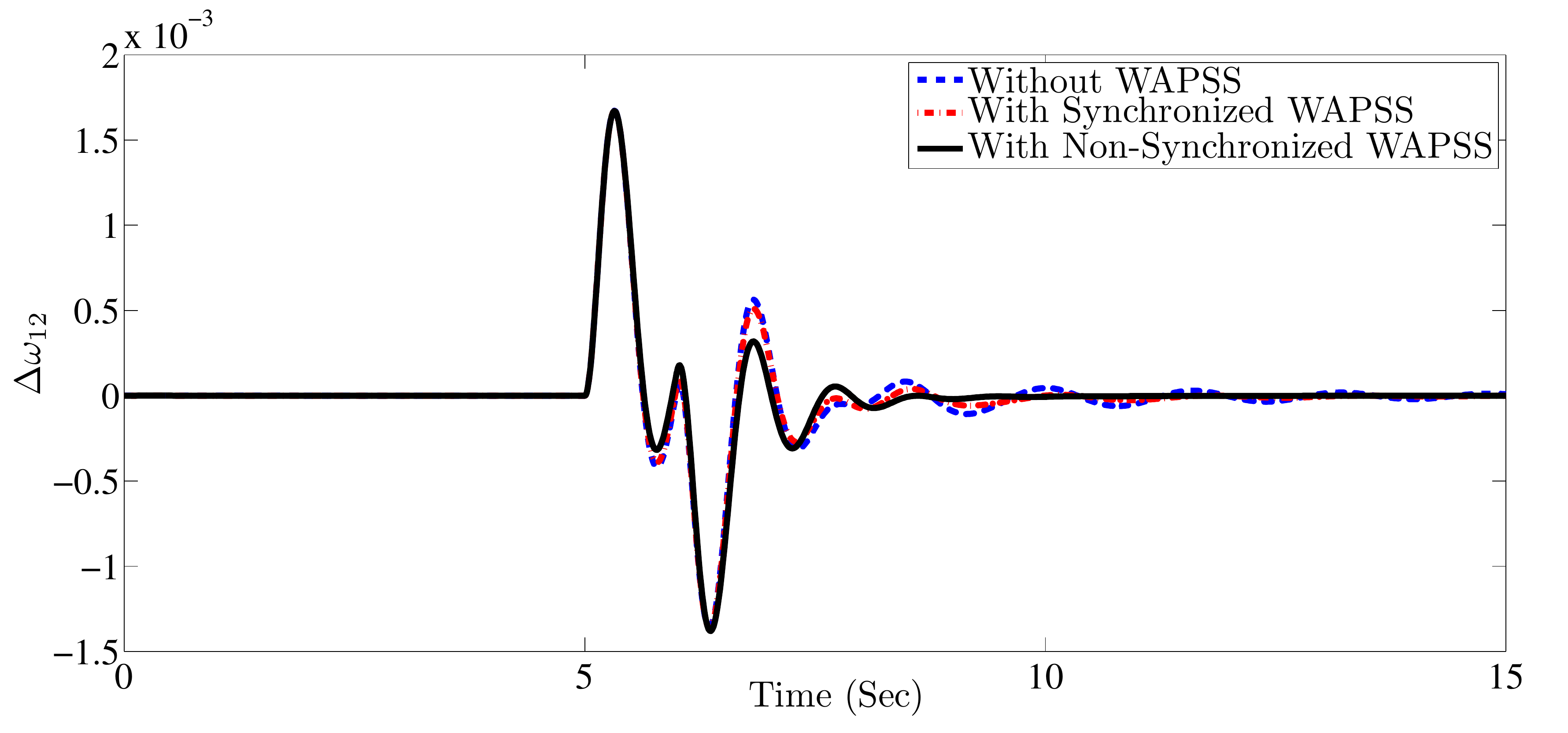}
    \label{fig:dW12_Syn_NonSyn}
    }
    \subfigure[]
    {%
     \includegraphics[width=3in,height=1.8in]{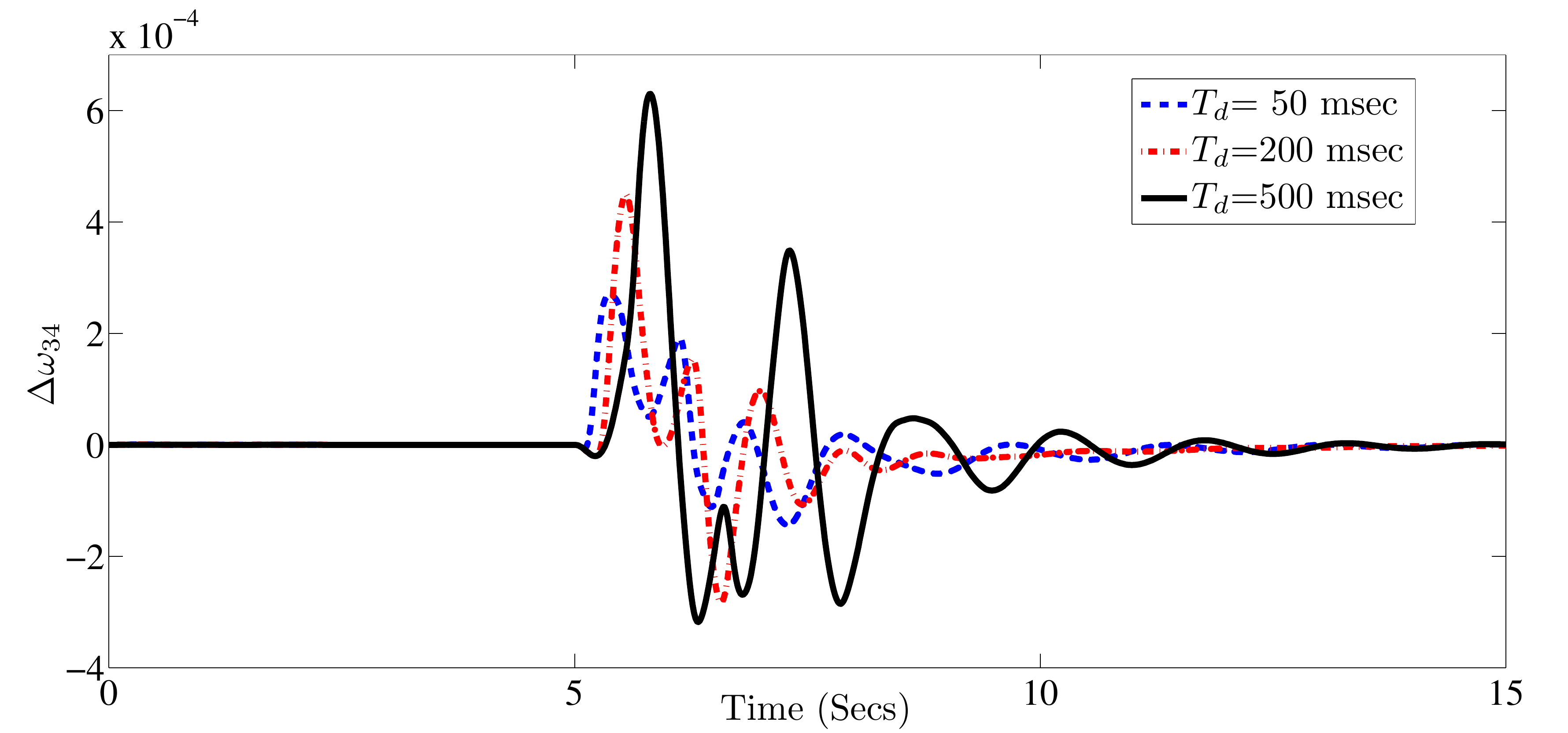}
     \label{fig:dW34_SynWAPSS_All}
    }
    \subfigure[]
    {%
     \includegraphics[width=3in,height=1.8in]{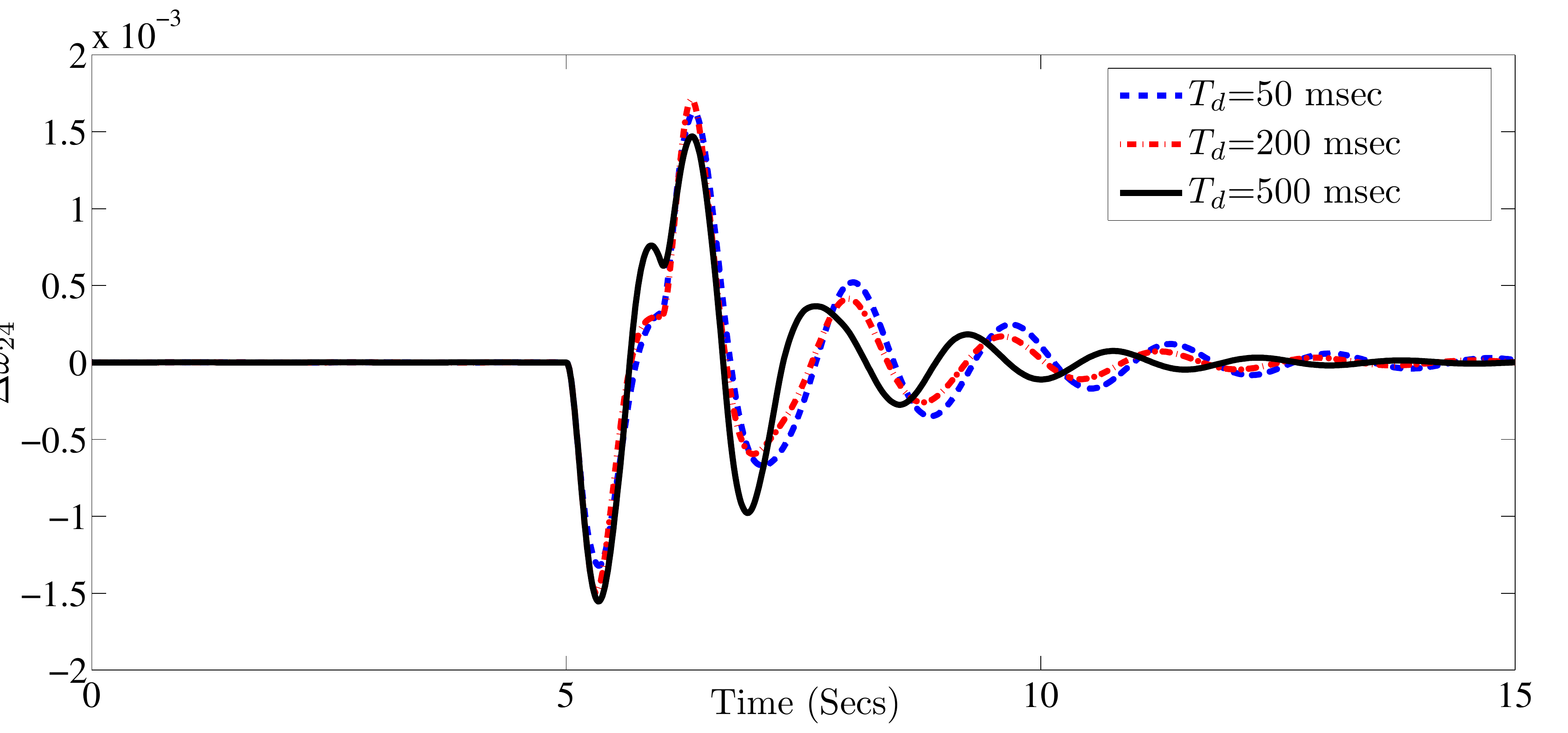}
     \label{fig:dW24_SynWAPSS_All}
    }
\caption{(a) $\Delta\omega_{24}$ variation:\color{blue}- - without WAPSS,\color{red}$-\cdot-$ with synchronized WAPSS,\color{black}---with non-synchronized WAPSS.(b) $\Delta\omega_{12}$ variation:\color{blue}- - without WAPSS,\color{red}$-\cdot-$ with synchronized WAPSS,\color{black}---with non-synchronized WAPSS. (c) $\Delta\omega_{34}$ variation with synchronized WAPSS for different delays ($T_d$). (d) $\Delta\omega_{24}$ variation with synchronized WAPSS for different delays ($T_d$).}
%  \label{fig_dW34vsGain}
\end{figure*}
\begin{figure*}[b]
\centering
    \subfigure[$\Delta\omega_{34}$ plot Non-synchronized WAPSS for multiple delays]
    {%
    \includegraphics[width=2.8in,height=1.85in]{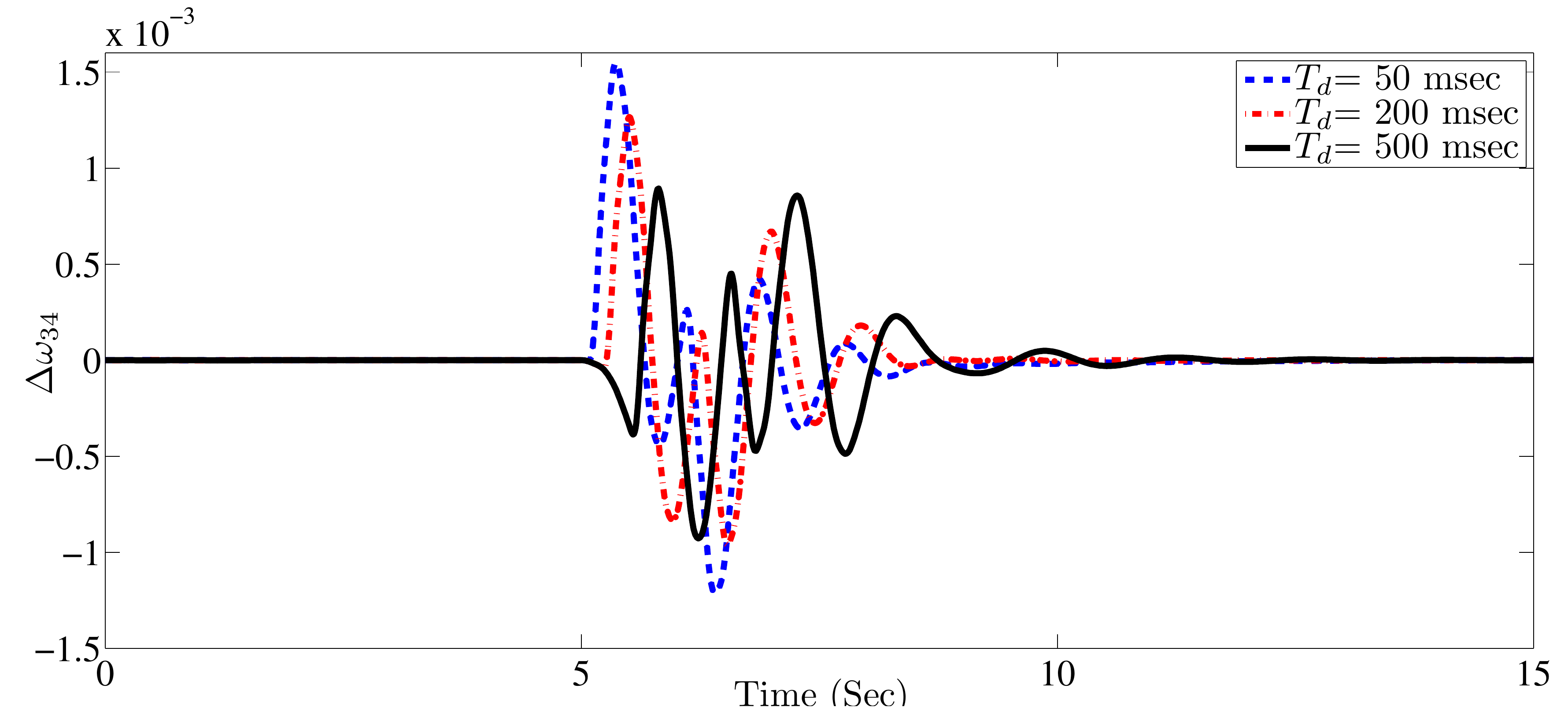}
        \label{fig:dW34_NonSynWAPSS_All}
    }
    \quad
    \subfigure[$\Delta\omega_{24}$ plot Non-synchronized WAPSS for multiple delays]
    {%
   \includegraphics[width=2.8in,height=1.85in]{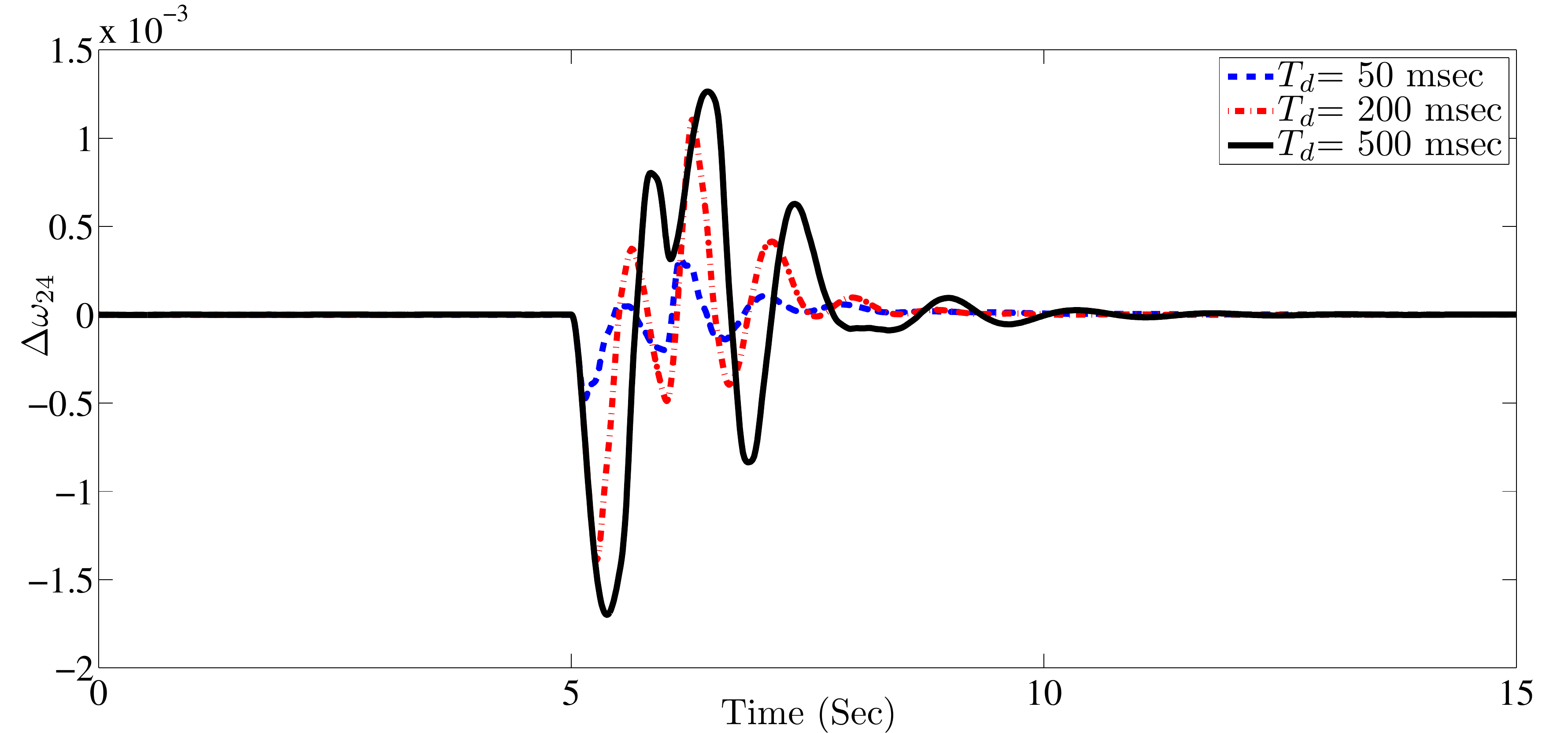}
        \label{fig:dW24_NonSynWAPSS_All}
    }
    \caption{Responses for Non-synchronized Controller with Multiple Delays}
  \label{fig:Case1-NonSynResponseswithDelays}
\end{figure*}
Now, considering the non-synchronized feedback where delay is present only in the remote signal, simulation responses are shown in Fig. \ref{fig:dW24_alldel_nonsyn}. It can be seen that the damping reduces as the delay is increased. However, as compared to the synchronized feedback, delay has lesser destabilizing effect.

%%%%%%%%%%%%%%%%%%%%%%%%%%%%%%%%%%%%%%%%%%%%%%%%%%%%%%%%%%%%%%%%%%%%%%%%%%%%%%%%%%%%%%%%%%%%%%%%%%%%%%%%%%%%%%%%%%%%%
\subsection{WAPSS Synthesis (With Considering Delay)}
From the above study, the importance of considering delay during the design stage is apparent. The controller is now designed considering a reasonable $200\ msec$ delay, following the procedure in Section-II. 
First considering synchronized feedback configuration, where an equal amount of delays are introduced in both the local and the remote signals. Responses to the disturbance are shown in Fig. \ref{fig:dW24_Syn_NonSyn}. It can be seen that oscillations died out within $10\ sec$ of disturbance occurrence.

Next, the controller is designed for non-synchronized feedback. From linear analysis, it is found out that the damping of inter-area mode has improved from $0.08$ to $0.255$. The speed deviation response to the disturbance is shown in Figs.\ref{fig:dW24_Syn_NonSyn}, \ref{fig:dW12_Syn_NonSyn}. The oscillation due to the inter-area mode dies out within 4 cycles of operation. The effect of WAPSS on the local mode of area-1 can be seen in Fig. \ref{fig:dW12_Syn_NonSyn}. The WAPSS still marginally improves its damping. 

To evaluate the performance, different delays are now considered for both the synchronized and the non-synchronized case. The controllers were designed considering $200\ msec$ time-delay and performances are evaluate for $50\ msec$, $200\ msec$ and $500\ msec$ time-delay. For synchronized feedback, though the system retains stability for different delays but performance degrades for the local mode. One can observe from Fig. \ref{fig:dW34_SynWAPSS_All} that oscillations take more time to be settled when the delay is other than $200\ msec$. However, the variation in delay has less significant effect on the inter-area modes as shown in Fig. \ref{fig:dW24_SynWAPSS_All}.

The performance of the non-synchronized WAPSS is also evaluated for the same time delays. It can be seen that the controller performs well for $500\ msec$ delay although the controller is designed for $200\ msec$ delay as shown in Fig.\ref{fig:dW24_NonSynWAPSS_All}. However, system performance improves if the delay is less than $200\ msec$, which is other way in case of the synchronized WAPSS. The delay in the remote signal, have the same effect on the local mode of area-2 as shown in Fig.\ref{fig:dW34_NonSynWAPSS_All}. Local oscillations are getting settled very quickly compared to the synchronized WAPSS. 

\begin{figure*}[!b]
\centering
\subfigure[]{
\includegraphics[width=3in,trim=0.2in 0 0.2in 0in,clip=true]{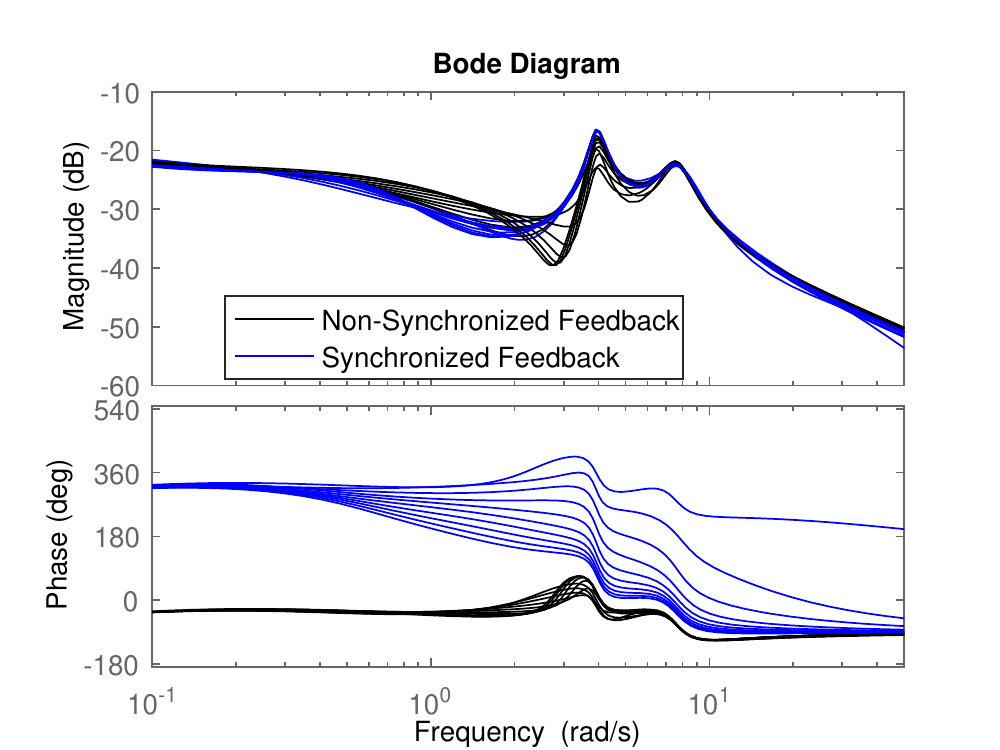}
}
\subfigure[]{
\includegraphics[width=3in,trim=0.2in 0 0.2in 0in,clip=true]{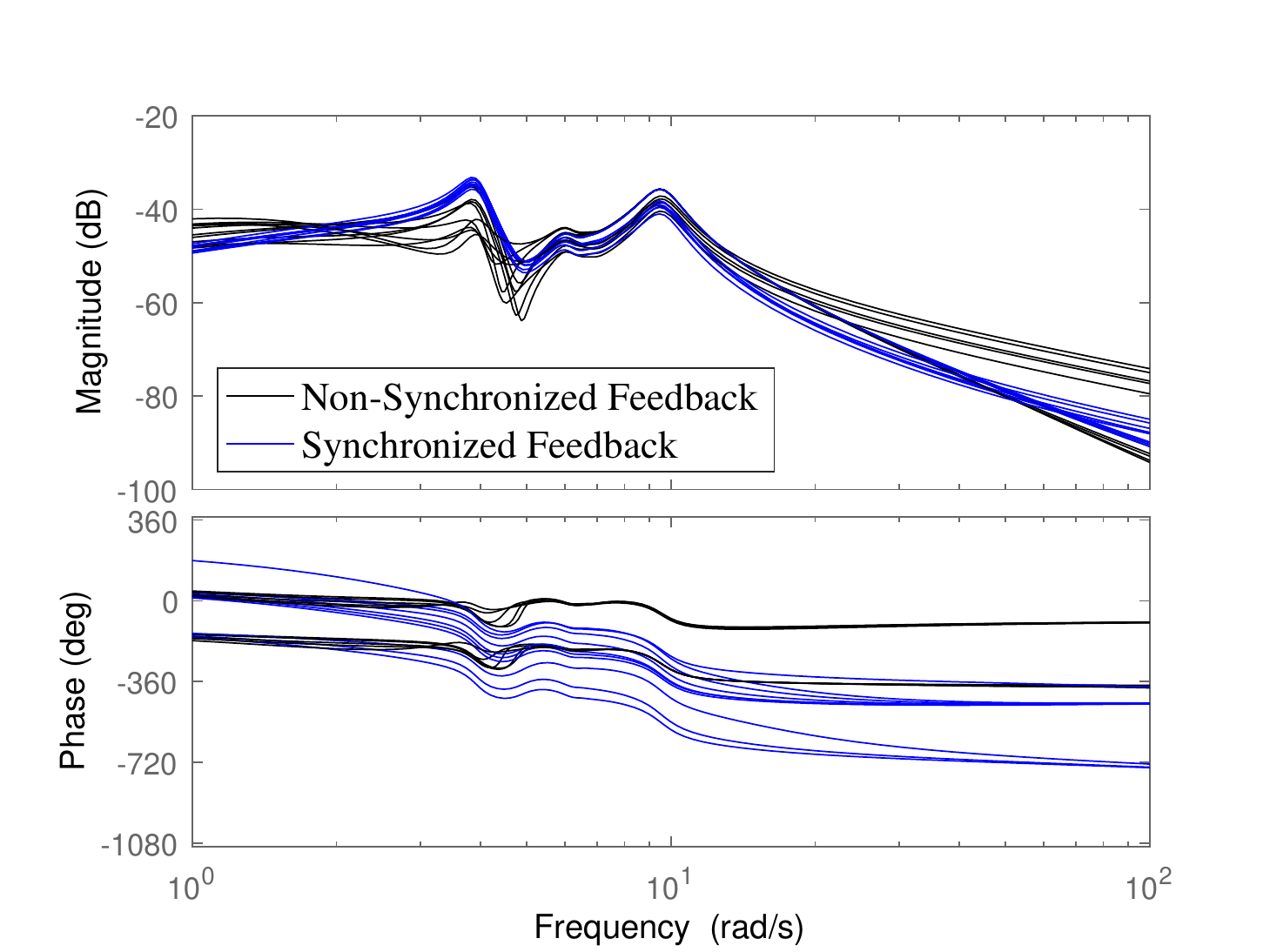}
}
\caption{Sensitivity of the closed-loop system for variations in delay value (a) for four machine and (b) for ten machine system.}
\label{Fig:sensitivity}
\end{figure*}

{
Another aspect of non-synchronized WAPSS is the reduced sensitivity to delay variations. For the system, if the delay is varied between $0.1\ sec$ to $1\ sec$, the ranges of the gain and phase for the sensitivity function (computed as (4), where $G_p(s)=G_{p0}(s)G_{d}(s)$, $G_{p0}(s)$ represents the nominal power system model and $G_{d}(s)$ is the delay approximated with second order Pade model) is much more for the synchronized feedback than the non-synchronized WAPSS as shown in Fig. \ref{Fig:sensitivity}. For designing damping controllers, phase compensation plays an important role, and in the case of the synchronized feedback with the delay variation the phase spectrum is more wider compared to the non-synchronized feedback which results in inferior performance for the synchronized WAPSS in the presence of the delay.}

\section{Case Study-II}
The 10 machine 39 bus benchmark system~\cite{pai89} is considered next with the local PSSs settings chosen as in ~\cite{jabr10}. The generators $G_1$ to $G_9$ are equipped with local PSSs. The nonlinear system model is constructed in MATLAB-Simulink and linearized around the operating point. From the modal analysis, it is found that three inter-area modes are present in the system as given in Table \ref{table_system_modes_ne} . The modes $\mathcal{M}_2$ and $\mathcal{M}_3$ have low damping but have relatively larger frequency compared to $\mathcal{M}_1$. Hence improving the damping of $\mathcal{M}_1$ ( $-0.28\pm j3.92$~) through WAPSS is only considered. Using geometrical measure, $\Delta\omega_{10,4}$ is chosen as the output and input of $G_4$ is selected for the WAPSS site. The linearized system model is of $96^{th}$-order, which leads to larger computation time and has implementation issue with the full-order $H_\infty$ controller. So the system model is reduced to a $6^{th}$ order one by means of Hankel norm reduction technique. Note that, the reduced order model retains the $\mathcal{M}_1$ mode.
\begin{figure}[!tbh]
\centering
\includegraphics[width=3.7in]{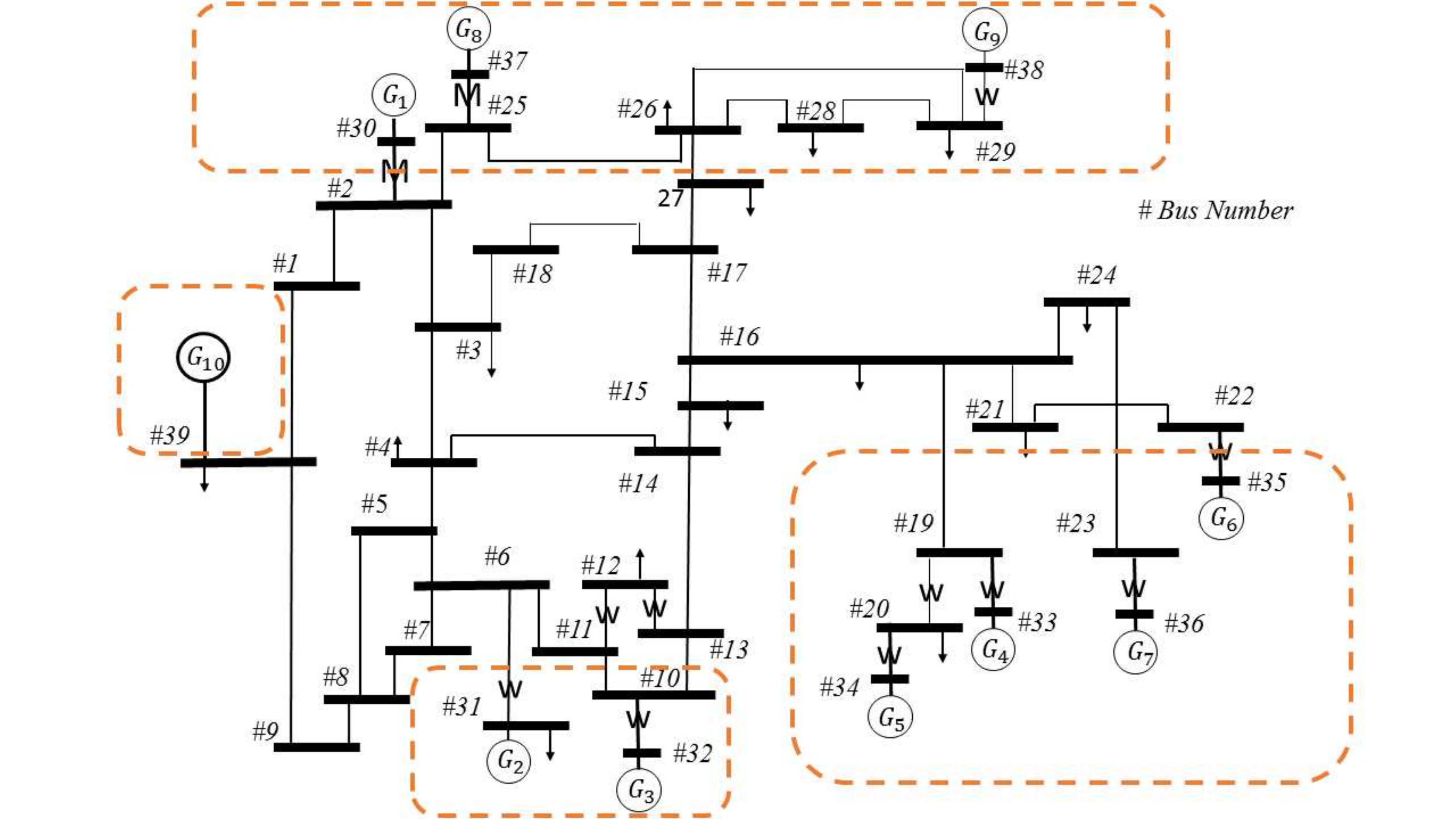}
 \caption{10 machine 39 bus (New England) system ~\cite{pai89}. }
\label{fig:10M39B}
\end{figure}

\begin{table}[!]
\renewcommand{\arraystretch}{1.3}
 \caption{Low-frequency modes of the system without WAPSS}
\label{table_system_modes_ne} \centering
\begin{tabular}{|c|c|c|c|}
\hline Mode & Mode Shape & Frequency (Hz) & Damping\\
\hline\hline $M_1$ & $G_{10}$ v/s $G_1-G_9$ & 0.6236 & 0.0705\\
\hline $M_2$ & $G_1,G_8,G_9$ v/s $G_4-G_8$ & 1.04 & 0.0703\\
\hline $M_3$ & $G_2,G_3$ v/s $G_4,G_5$ & 0.96 & 0.0691\\
\hline
\end{tabular}
\end{table}
\subsection{WAPSS Synthesis (Without Considering Delay)}
\begin{figure*}[!]
\centering
    \subfigure[]
    {%
     \includegraphics[width=3in,height=1.8in]{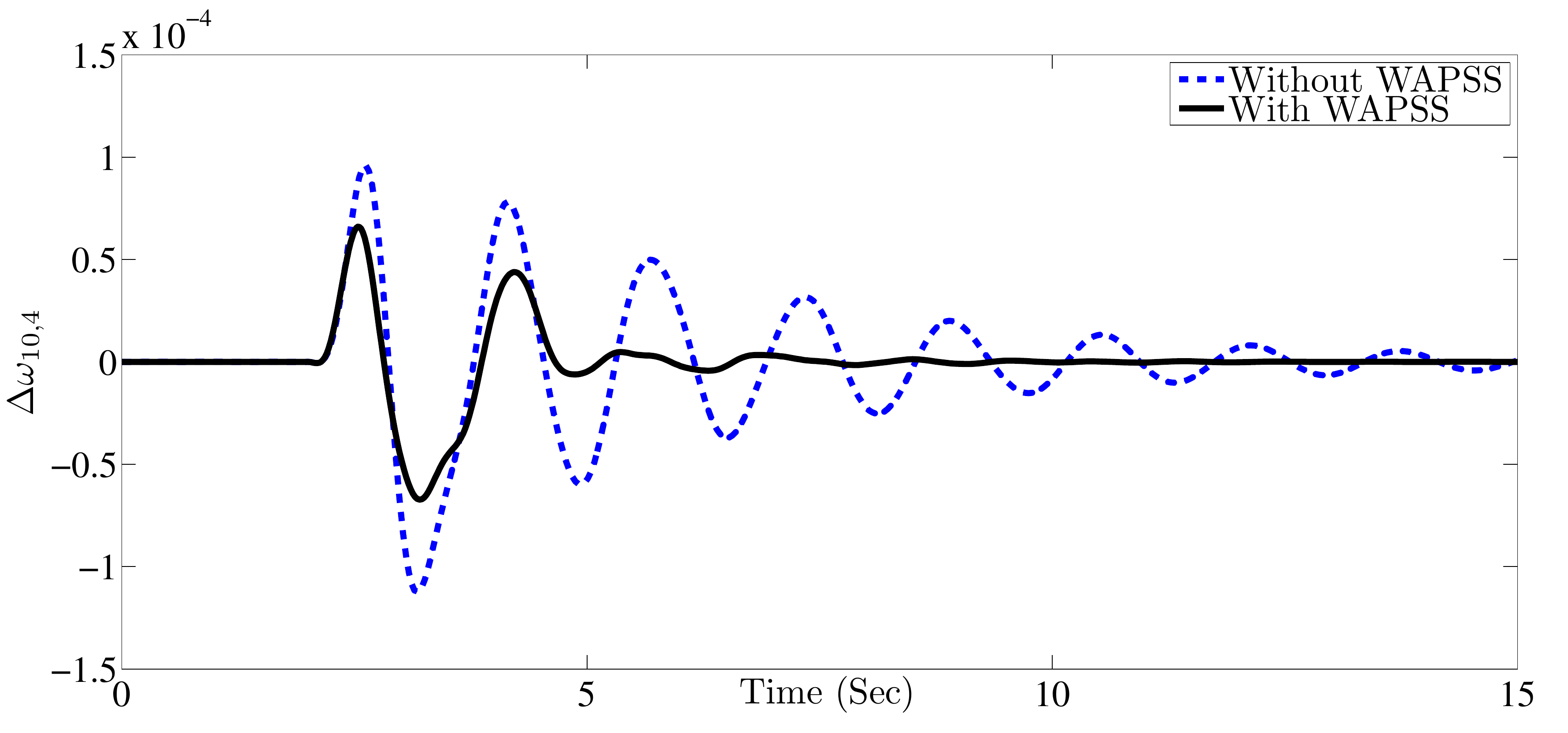}
     \label{fig:dW104_WAPSS}
    }
    \subfigure[]
    {%
    \includegraphics[width=3in,height=1.8in]{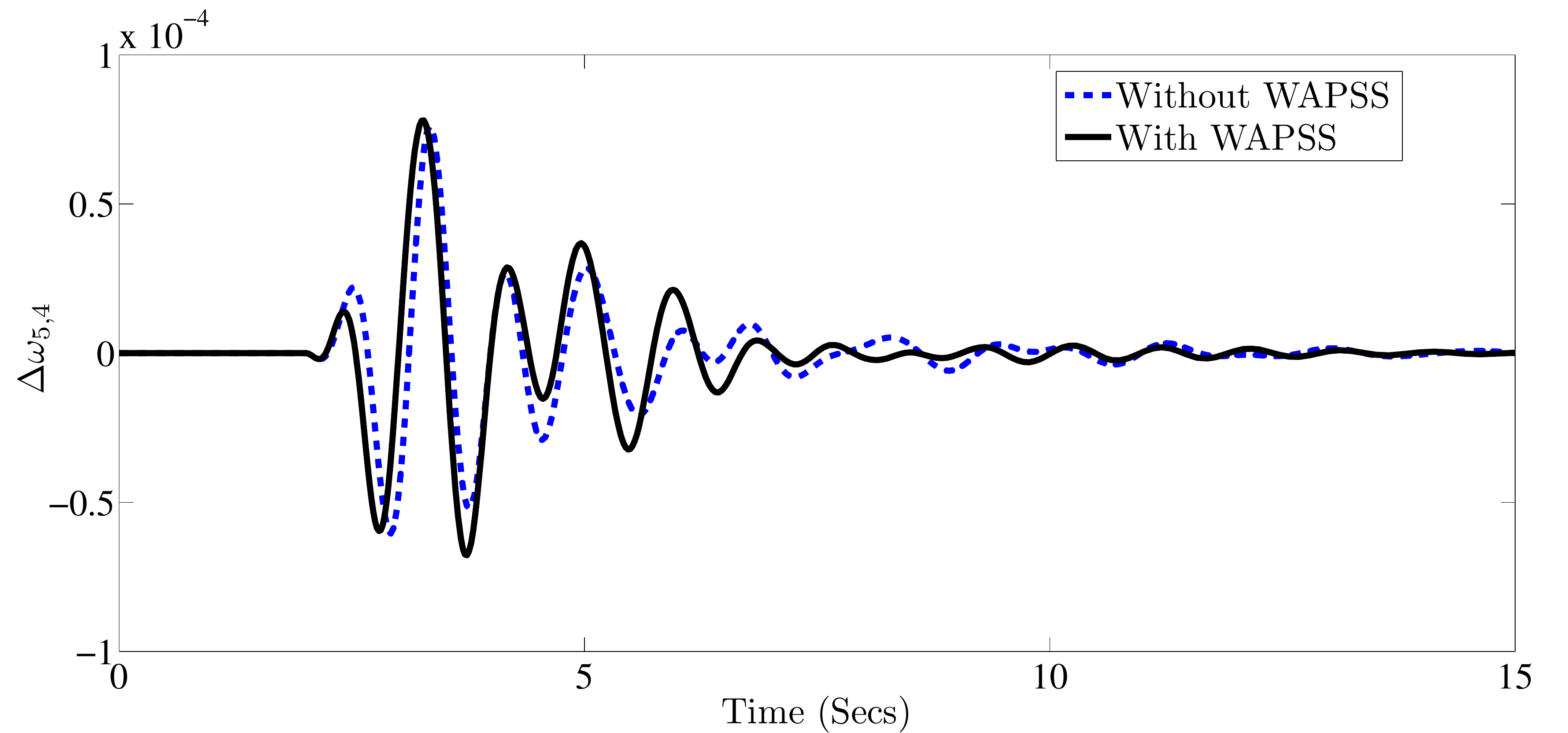}
\label{fig:dW54_WAPSS}
    }
    \subfigure[]
    {%
\includegraphics[width=3in,height=1.8in]{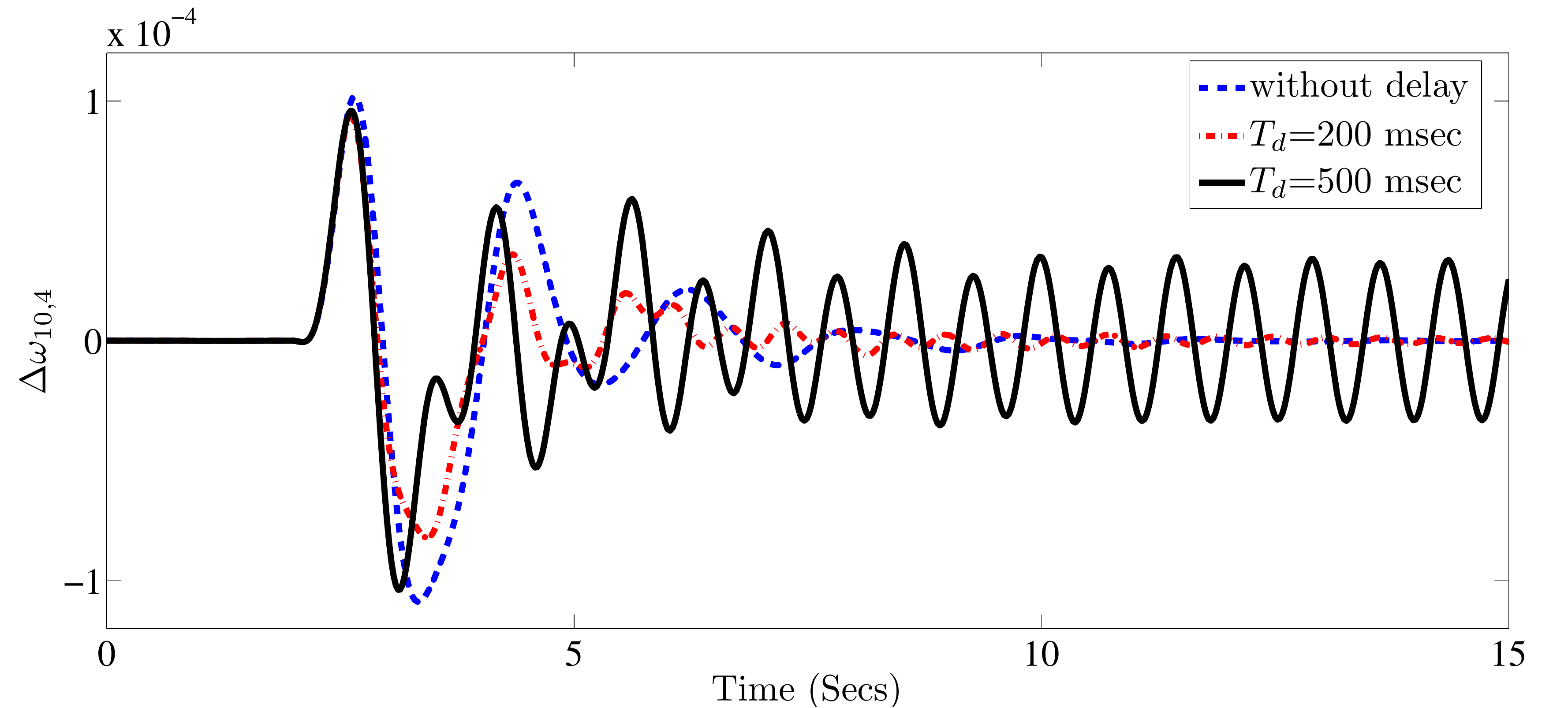}
\label{fig:CS2_Synchronization_Delay}
    }
    \subfigure[]
    {%
\includegraphics[width=3in,height=1.8in]{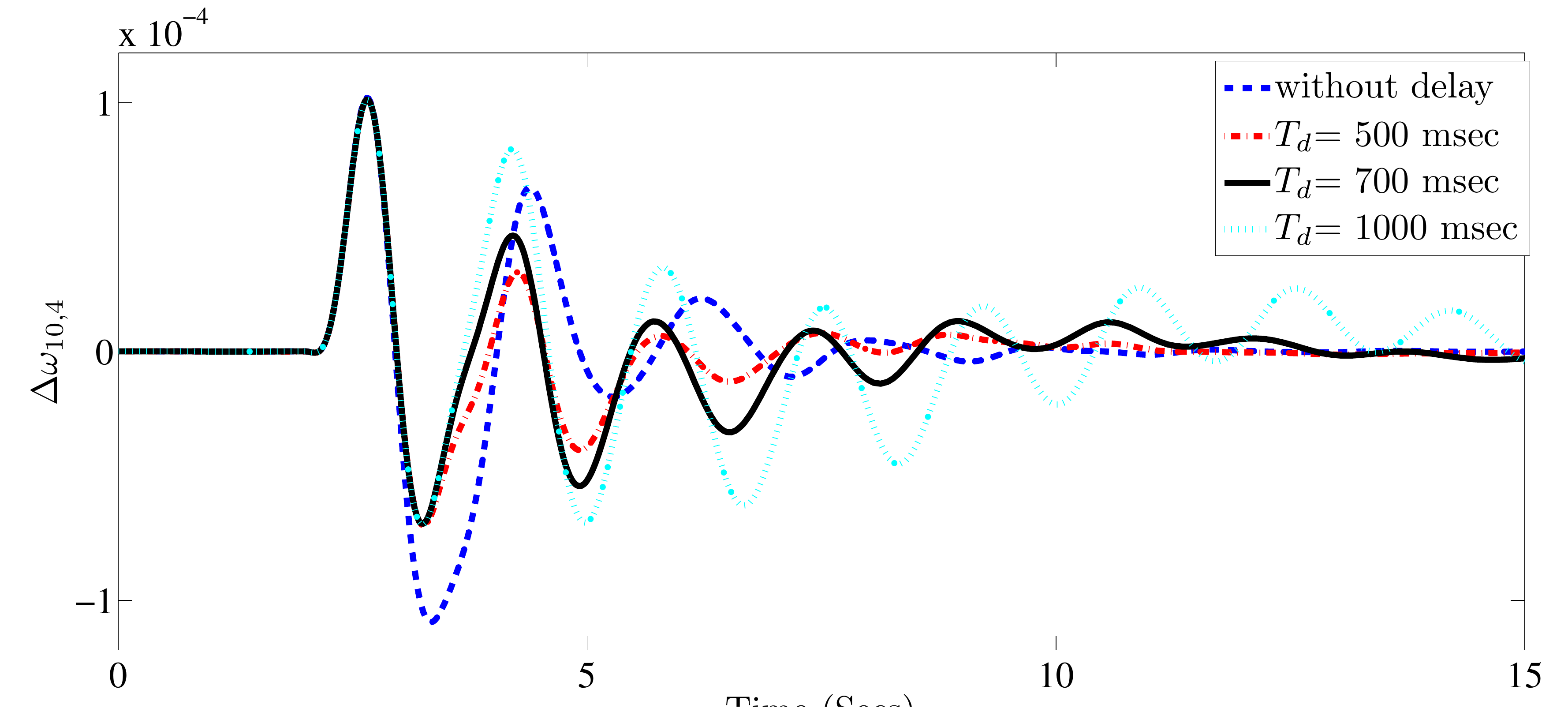}
\label{fig:CS2_Non_Synchronization_Delay}
    }
\caption{(a) $\Delta\omega_{10,4}$ variation: - - without WAPSS,--- with WAPSS (b) $\Delta\omega_{5,4}$ variation:- - without the WAPSS,--- with the WAPSS.(c) $\Delta\omega_{10,4}$ variation for different delays ($T_d$) in synchronized feedback.(d) $\Delta\omega_{10,4}$ variation for different delays ($T_d$) in non-synchronized feedback.}
%  \label{fig_dW34vsGain}
\end{figure*}

The WAPSS is first designed without considering the delay, as in section-II. With the designed WAPSS, the closed-loop location of the inter-area mode $\mathcal{M}_1$ is found to be $-1.10\pm j3.91$. Therefore, damping of the mode is improved from $0.07$ to $0.271$. Also, the performance of the WAPSS is evaluated through nonlinear simulation with a $0.2\ sec$-pulse disturbance of $0.05$ magnitude applied at the $G_4$ terminal's voltage reference. As inter-area oscillation can be observed in the difference of speed deviation between $G_{10}$ and $G_4$, the response of this is shown in Fig. \ref{fig:dW104_WAPSS}. It can be seen that the oscillations are damped within the desired time compared to the case without the wide-area controller. Whereas, the WAPSS has no counter effect on the local modes as shown in Fig. \ref{fig:dW54_WAPSS}.

%\begin{figure}[!tbh]
%\centering
%\includegraphics[width=3.5in,height=2in]{Figures/dW_104_WAPSS.pdf}
% \caption{$\Delta\omega_{10,4}$ variation: - - without WAPSS,--- with WAPSS.}
%\label{fig:dW104_WAPSS}
%\end{figure}
%
%\begin{figure}[!tbh]
%\centering
%\includegraphics[width=3.5in,height=2in]{Figures/dW_54_WAPSS.pdf}
%\caption{$\Delta\omega_{5,4}$ variation:- - without the WAPSS,--- with the WAPSS.}
%\label{fig:dW54_WAPSS}
%\end{figure}
To study the effect of delay in the wide-area loop, a set of delays are now considered. The effect of delay in the synchronized configuration can be seen in Fig. \ref{fig:CS2_Synchronization_Delay}. The damping becomes poorer with increase in delay and finally becomes unstable with delay more than $500\ msec$. Next, delay is considered in the non-synchronized configuration and the corresponding response is shown in Fig. \ref{fig:CS2_Non_Synchronization_Delay}. It can be seen that the non-synchronized configuration can tolerate more delay compared with the synchronized case. It can be seen that even when the delay was increased up to $1\ sec$, the system was still stable. However, damping reduces with reduces with increase in delay. From the study, it can be concluded that the effect of delay has to be considered while designing the controller.
%\begin{figure}[!tbh]
%\centering
%\includegraphics[width=3.5in,height=2in]{Figures/dW_104_SynDelayEffect.pdf}
% \caption{$\Delta\omega_{10,4}$ variation for different delays ($T_d$) in synchronized feedback.}
%\label{fig:CS2_Synchronization_Delay}
%\end{figure}
%
%\begin{figure}[!tbh]
%\centering
%\includegraphics[width=3.5in,height=2in]{Figures/dW_104_NonSynDelayEffect.pdf}
% \caption{$\Delta\omega_{10,4}$ variation for different delays ($T_d$) in non-synchronized feedback.}
%\label{fig:CS2_Non_Synchronization_Delay}
%\end{figure}

\subsection{WAPSS Synthesis (With Considering Delay)}
\begin{figure*}[!]
\centering
    \subfigure[]
    {%
      \includegraphics[width=3in,height=1.8in]{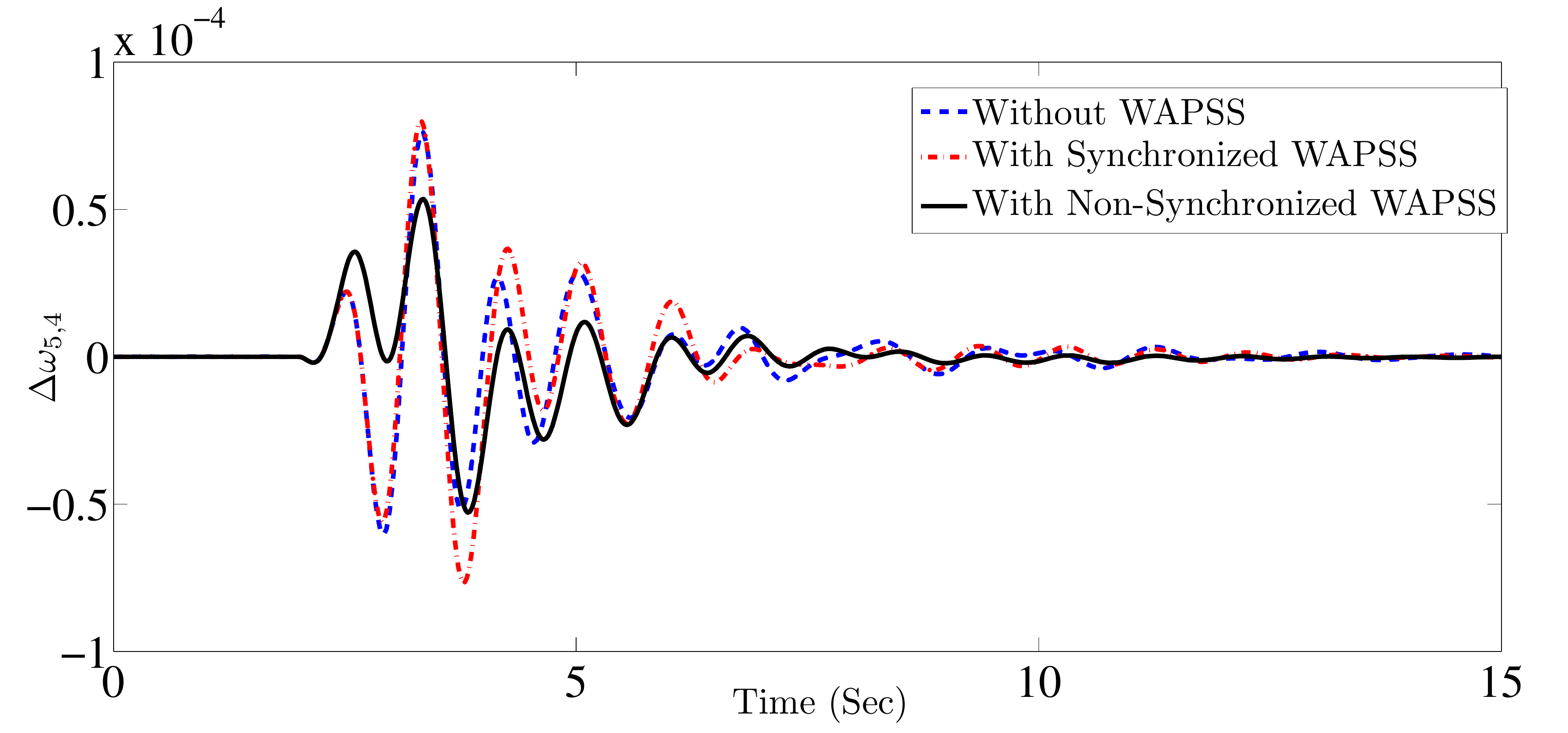}
      \label{fig:dW54_Syn_NonSyn}
    }
    \subfigure[]
    {%
\includegraphics[width=3in,height=1.8in]{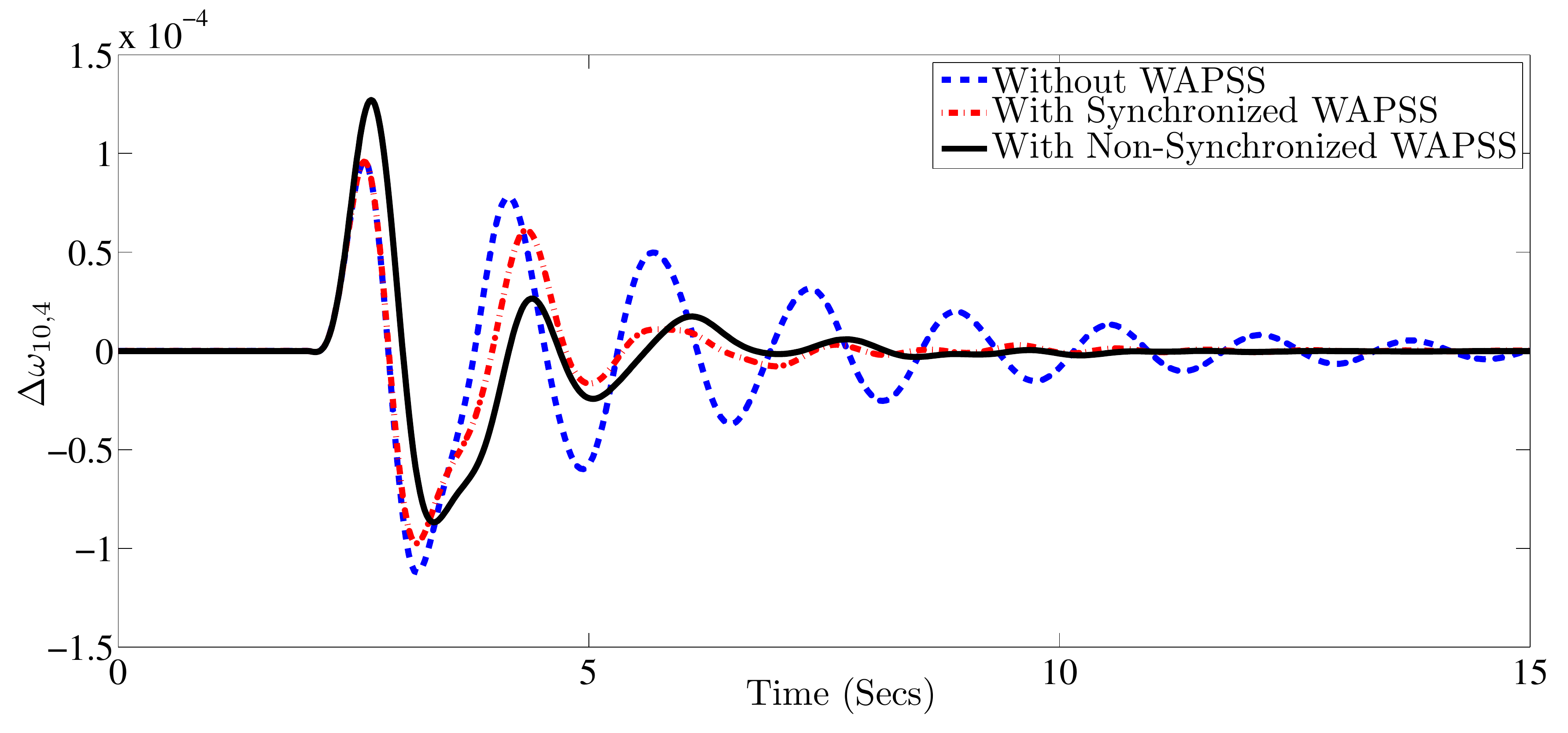}
\label{fig:dW104_Syn_NonSyn}
    }
    
    \subfigure[]
    {%
\includegraphics[width=3in,height=1.8in]{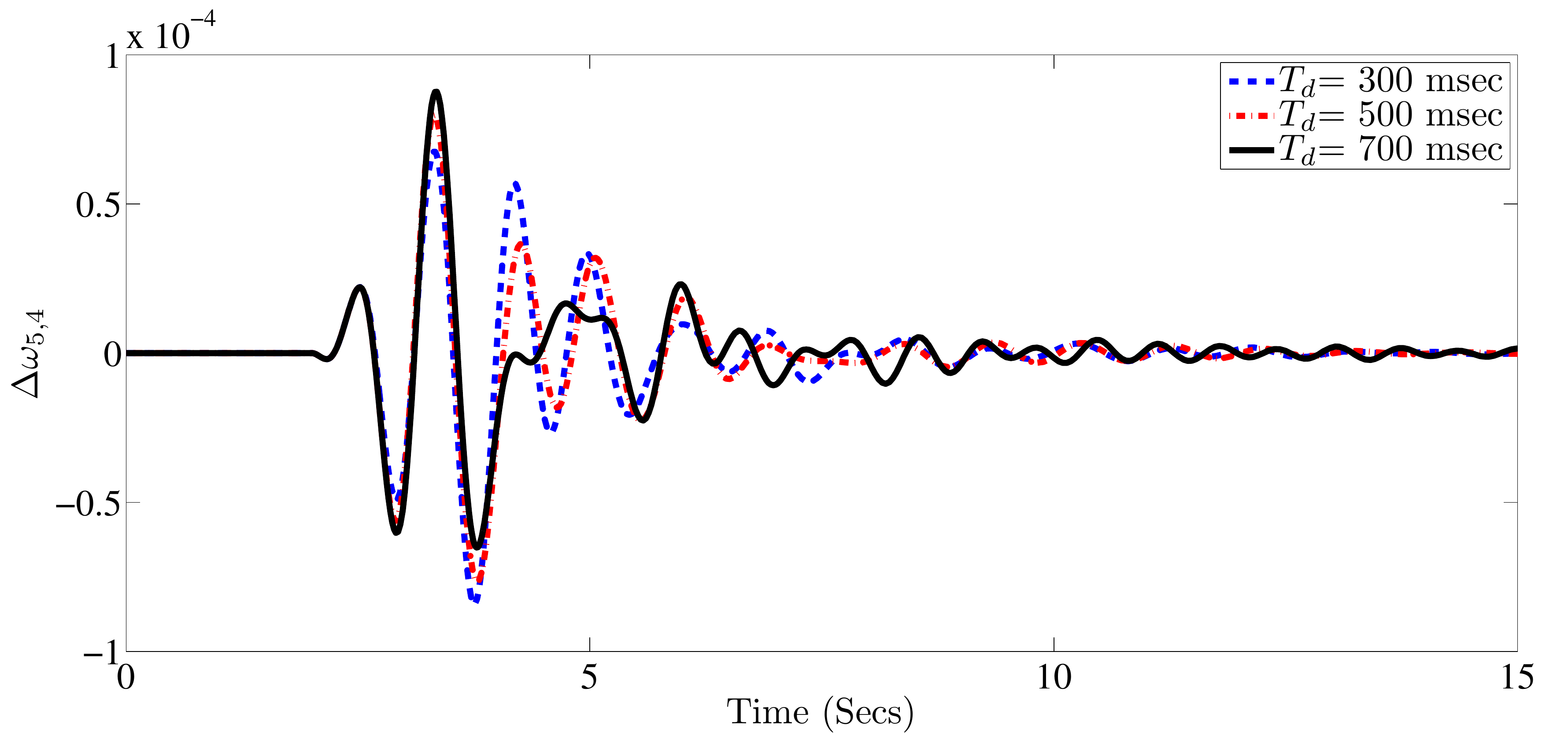}
\label{fig:dW54_SynWAPSS_All}
    }
    \subfigure[]
    {%
\includegraphics[width=3in,height=1.8in]{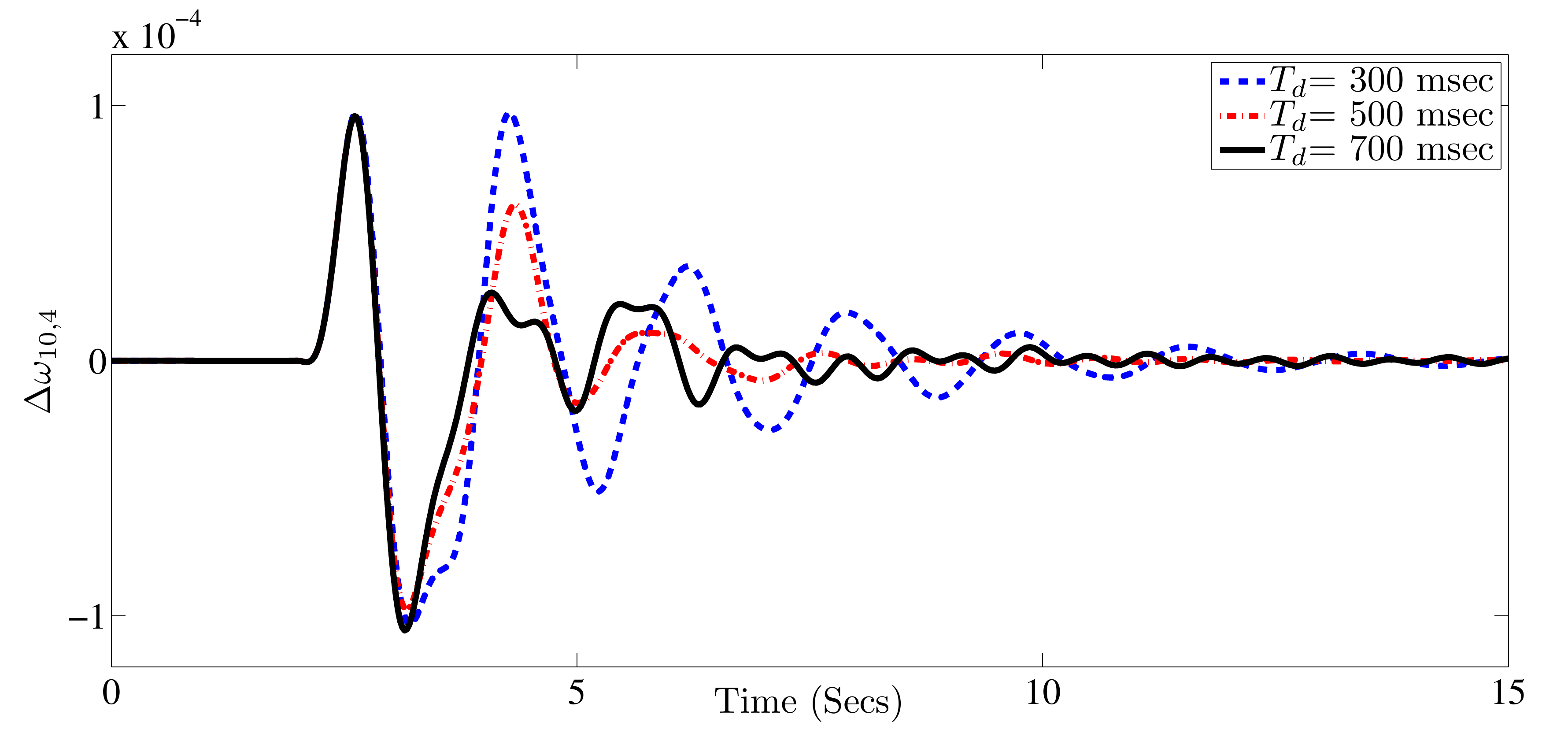}
\label{fig:dW104_SynWAPSS_All}
    }
\caption{(a) $\Delta\omega_{5,4}$ variation:\color{blue}- - without WAPSS,\color{red}$-\cdot-$ with synchronized WAPSS,\color{black}---with non-synchronized WAPSS. (b) $\Delta\omega_{10,4}$ variation:\color{blue}- - without WAPSS,\color{red}$-\cdot-$ with synchronized WAPSS,\color{black}---with non-synchronized WAPSS. (c) $\Delta\omega_{5,4}$ variation for synchronized WAPSS for different delays ($T_d$). (d) $\Delta\omega_{10,4}$ variation for synchronized WAPSS for different delays ($T_d$).}
%  \label{fig_dW34vsGain}
\end{figure*}

The WAPSSs are designed considering a delay of $500\ msec$ delay in synchronized and non-synchronized feedback. With the synchronized WAPSS, the inter-area mode is found to be $-0.942 \pm j3.87$. Therefore, the the damping is increased from $0.07$ to $0.236$. The performance of the designed controller is validated in nonlinear simulation where disturbance is added through $G_4$ as $0.2$ sec pulse disturbance. The variation in speed deviation difference is shown in Fig. \ref{fig:dW54_Syn_NonSyn} and Fig. \ref{fig:dW104_Syn_NonSyn}, where it can be seen that the oscillations are settled down quickly. Next, with non-synchronized WAPSS the inter-area mode is found out to be $-1.03\pm j3.64$. Therefore, damping of the inter-area mode is improved from $0.07$ to $0.273$. A disturbance is added as in the previous case to validate the controller performance. The responses to the disturbance in terms of speed deviation difference are shown in Fig. \ref{fig:dW54_Syn_NonSyn} and Fig. \ref{fig:dW104_Syn_NonSyn} along with synchronized feedback. One can observe that with non-synchronized WAPSS, oscillations of local mode and also inter-area mode is settled with in $10 sec$. Therefore, the non-synchronized WAPSS performs better than the synchronized one

%\begin{figure}[!tbh]
%\centering
%\includegraphics[width=3.5in,height=2in]{Figures/dW_54_Syn_NonSyn_WAPSS.pdf}
% \caption{$\Delta\omega_{5,4}$ variation:\color{blue}- - without WAPSS,\color{red}$-\cdot-$ with synchronized WAPSS,\color{black}---with non-synchronized WAPSS.}
%\label{fig:dW54_Syn_NonSyn}
%\end{figure}
%
%\begin{figure}[!tbh]
%\centering
%\includegraphics[width=3.5in,height=2in]{Figures/dW_104_Syn_NonSyn_WAPSS.pdf}
% \caption{$\Delta\omega_{10,4}$ variation:\color{blue}- - without WAPSS,\color{red}$-\cdot-$ with synchronized WAPSS,\color{black}---with non-synchronized WAPSS.}
%\label{fig:dW104_Syn_NonSyn}
%\end{figure}

%\begin{figure}[!tbh]
%\centering
%\includegraphics[width=3.5in,height=2in]{Figures/dW_54_NonSynWAPSS.pdf}
% \caption{$\Delta\omega_{5,4}$ plot for Non-synchronized WAPSS}
%\label{fig:dW54_NonSynWAPSS}
%\end{figure}
%
%\begin{figure}[!tbh]
%\centering
%\includegraphics[width=3.5in,height=2in]{Figures/dW_104_NonSynWAPSS.pdf}
% \caption{$\Delta\omega_{10,4}$ plot for Non-synchronized WAPSS}
%\label{fig:dW104_NonSynWAPSS}
%\end{figure}

\begin{figure*}[!]
\centering
    \subfigure[]
    {%
     \includegraphics[width=3in,height=1.8in]{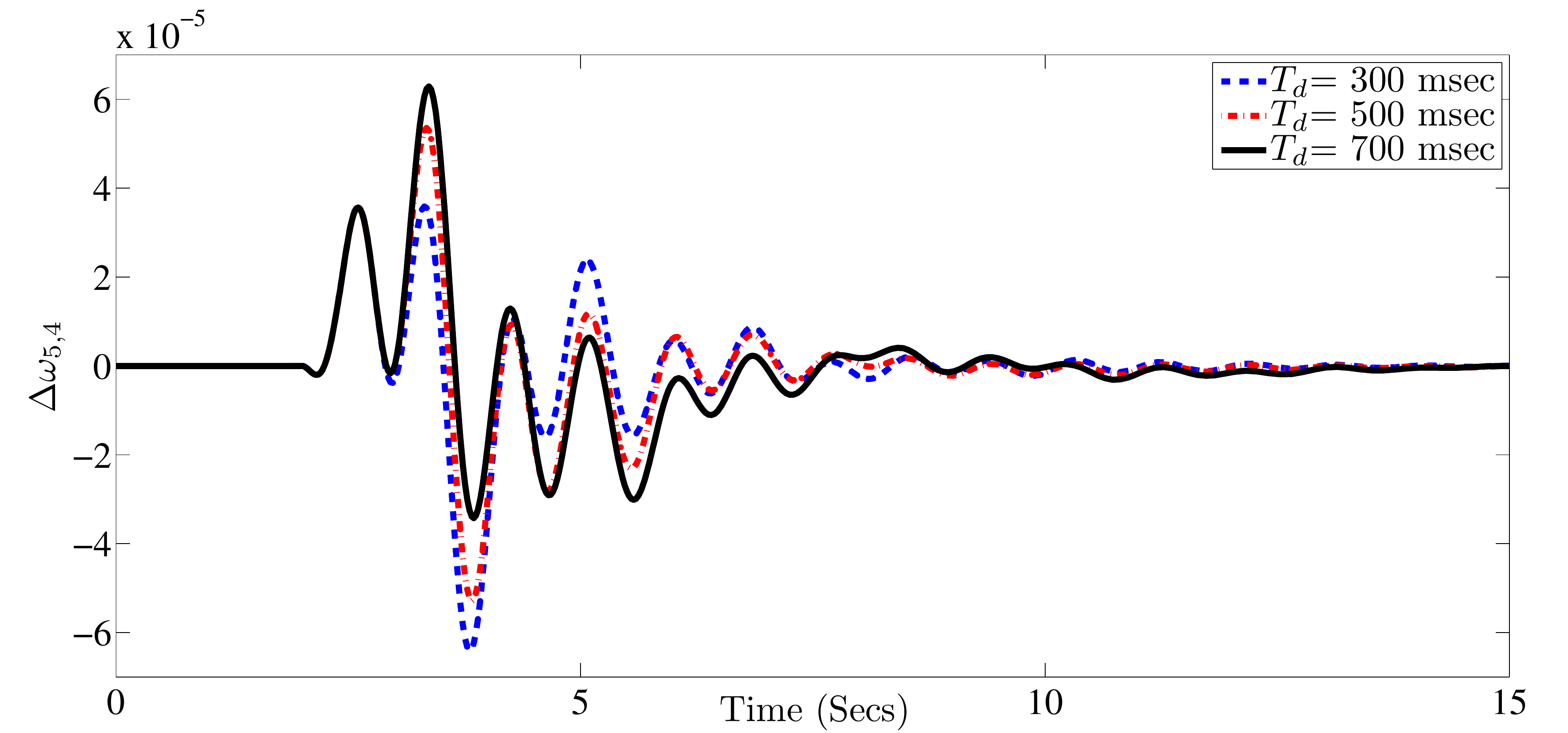}
\label{fig:dW54_NonSynWAPSS_All}
    }
    \subfigure[]
    {%
\includegraphics[width=3in,height=1.8in]{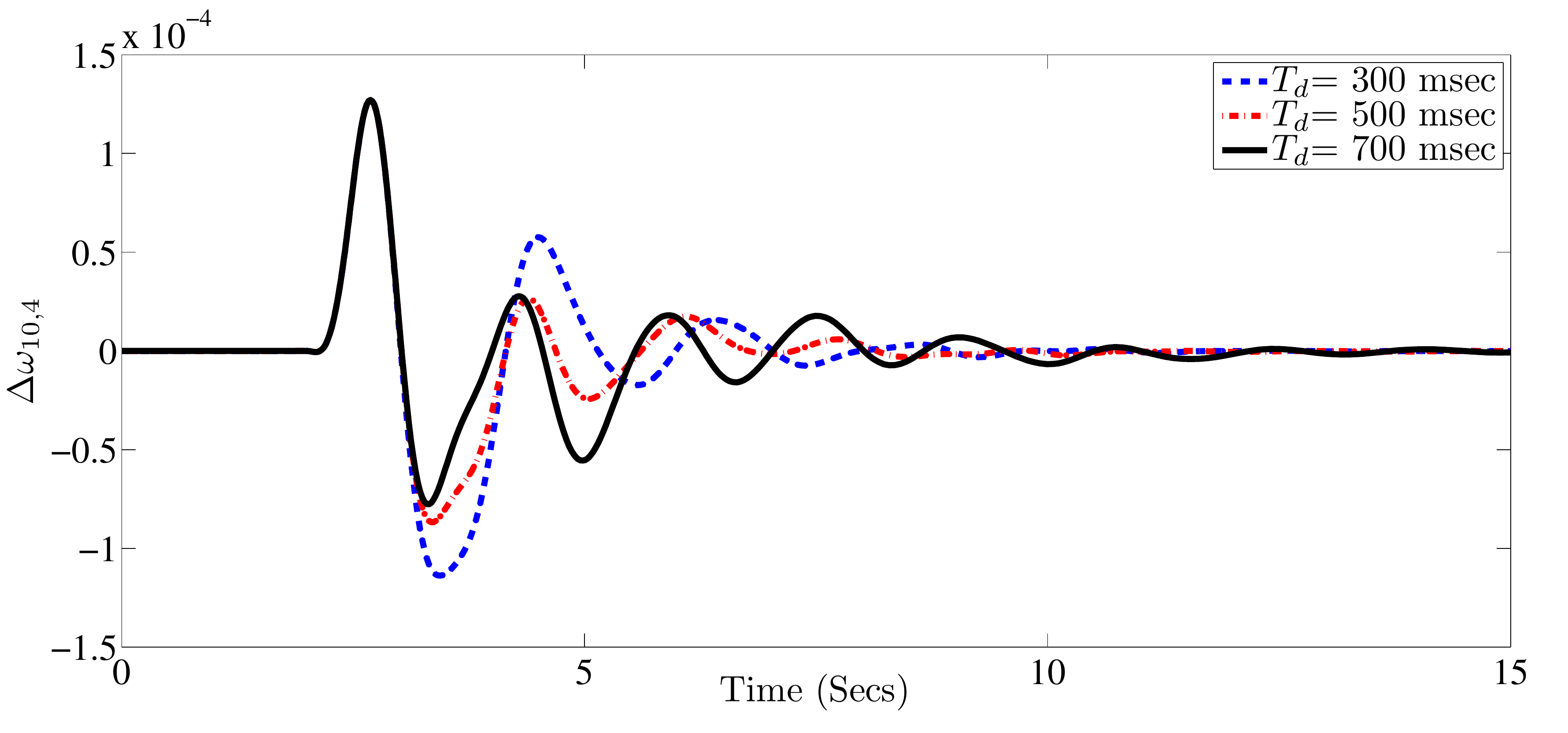}
\label{fig:dW104_NonSynWAPSS_All}
    }
  \caption{(a) $\Delta\omega_{5,4}$ variation for non-synchronized WAPSS for different delays (b) $\Delta\omega_{10,4}$ variation with non-synchronized WAPSS for different delays.}
%  \label{fig_dW34vsGain}
\end{figure*}

%\subsection{Evaluation of the Performances}
Next, to evaluate the performance of the synchronized WAPSS and the non-synchronized WAPSS, different delays are considered. As the controller is designed with $500\ msec$ delay, so one more than it ($700\ msec$)  another lesser ($300\ msec$) are considered. The simulation result with these delays are shown in Figs. \ref{fig:dW54_SynWAPSS_All}, \ref{fig:dW104_SynWAPSS_All}, \ref{fig:dW104_NonSynWAPSS_All}, \ref{fig:dW54_NonSynWAPSS_All}. 

It can be seen that with increase in delay, the damping reduces but the effect is more prominent in local modes. With synchronized WAPSS, though the system retains stability for both the change in delay, performance degrades significantly. For the delay less than the value used during the design, settling time has increased but it stays less than $10\ sec$ and for the delay larger than $500\ msec$ a faster mode of low magnitude has been observed in speed deviations. The same effect has been observed in inter-area oscillations also. For non-synchronized case, no such effect has been observed. As the delay is not present in local signal, the change in delay has no significant effect on the local mode. On the other side, for delay less than $500\ msec$ the system performance is better with inter-area oscillations. Similar to case-I, sensitivity is studied for the variations in delay as in Fig. \ref{Fig:sensitivity}. A similar trend as in Case-I,  where synchronized feedback is more sensitive to delay variations compared to non-synchronized WAPSS, is observed.
%
%\begin{figure}[!tbh]
%\centering
%\includegraphics[width=3.5in,height=2in]{Figures/dW_54_SynWAPSS_DiffDelay.pdf}
% \caption{$\Delta\omega_{5,4}$ variation for synchronized WAPSS for different delays ($T_d$).}
%\label{fig:dW54_SynWAPSS_All}
%\end{figure}
%
%\begin{figure}[!tbh]
%\centering
%\includegraphics[width=3.5in,height=2in]{Figures/dW_104_SynWAPSS_DiffDelay.pdf}
% \caption{$\Delta\omega_{10,4}$ variation for synchronized WAPSS for different delays ($T_d$).}
%\label{fig:dW104_SynWAPSS_All}
%\end{figure}

%\begin{figure}[!tbh]
%\centering
%\includegraphics[width=3.5in,height=2in]{Figures/dW_54_NonSynWAPSS_DiffDelay.pdf}
% \caption{$\Delta\omega_{5,4}$ variation for non-synchronized WAPSS for different delays}
%\label{fig:dW54_NonSynWAPSS_All}
%\end{figure}
%
%\begin{figure}[!tbh]
%\centering
%\includegraphics[width=3.5in,height=2in]{Figures/dW_104_NonSynWAPSS_DiffDelay.pdf}
% \caption{$\Delta\omega_{10,4}$ variation with non-synchronized WAPSS for different delays.}
%\label{fig:dW104_NonSynWAPSS_All}
%\end{figure}

%\section{Observation}
%It has been observed that with local PSS, damping of the inter-area mode is poor, which can be improved with WAPSS. However, if the delay is not considered during the design stage, the introduction of delay in the wide-area signal during the operation of the controller reduces the damping or may cause instability. The delay effects differently depending on synchronization of signals, it is observed that a non-synchronized feedback are more tolerable to delay than a synchronized feedback. A $2^{nd}$ order Pade approximated model works well for synthesis of the controller. A non-synchronized WAPSS performs better compared with synchronized WAPSS. Also, the controller designed with non-synchronized feedback are more robust towards the delay variations specially to local modes. One can switch of the wide-area signal off and on without hampering stability of the system for non-synchronized case, whereas this is not possible for synchronized case. The above designed controller shows effective results in analysis and simulation.
\section{Conclusion}
{Damping improvement of inter-area modes through wide-area controller, particularly through WAPSS, has been discussed in this paper. For such WAPSS, when speed  information is used, the feedback configuration can be structured in synchronized or non-synchronized fashion. Considering these two configurations at the design stage, performance of the designed controllers have been investigated in this paper. It has been shown that the design controllers are effective to exploit the benefits of the non-synchronized configuration shown in earlier work \cite{ghosh16}. The WAPSS has been designed in the framework of $H_\infty$ control with regional pole placement. The effect of delay on the performance of the WAPSS has been investigated while considering the $2^{nd}$ order Pade approximation of the delay in the design procedure. The effectiveness of the controller has been validated with two case studies. It has been shown the the non-synchronized WAPSS performed better compared to the synchronized WAPSS while considering robustness towards delay variations.}
%\section{Acknowledgment}

%

%\clearpage

\bibliography{ReferencesWAC}
\bibliographystyle{IEEEtran}

\section*{Appendices}%$\appendices
\section{Proof of Theorem-1~\cite{chilali96}}
Theorem 1 presented in this work follows the results of \cite{chilali96}. For completeness, a brief proof of the same is presented here. More details can be found in the original work. Consider a system as in (\ref{eq:plant}) and the controller to be (\ref{eq:controller}) resulted closed-loop system in the form:
\begin{equation}\label{eq:cl_appe}
\begin{split} 
\dot{x}&=Ax+Bw,\\
z&=Cx+Dw,
\end{split}
\end{equation}
where
\begin{equation*}
A=\begin{bmatrix}
A_p&B_pC_c\\
B_cC_p&A_c
\end{bmatrix}
B=\begin{bmatrix}
B_w\\
0
\end{bmatrix}
C=\begin{bmatrix}
C_z&0
\end{bmatrix}
D=D_z.
\end{equation*}
Considering the $H_\infty$ norm as performance measure of the system (\ref{eq:cl_appe}),
\begin{equation}\label{Hinf_LMI} 
\begin{bmatrix}
    A^TP+PA&PB&C^T \\
    B^TP & -\gamma I & D^T \\
    C &D&-\gamma I  \\
\end{bmatrix}
  < 0,
\end{equation}
and the complexity arises with bilinearity in $A^TP+PA$ or $PB$.

Let $P=\begin{bmatrix}
S&V\\
V^T&*\\
\end{bmatrix}$
and  $P^{-1}=\begin{bmatrix}
R&U\\
U^T&*\\
\end{bmatrix},$
where $S$ and $R$ are symmetric matrix of appropriate dimensions. As $PP^{-1}=I$, so it can be said as
 $$P\left({\begin{array}{cc} R&I\\U^T&0\end{array}}\right)=\left({\begin{array}{cc} I&S\\0&V^T\end{array}}\right),$$ also can be written as:
\begin{equation*}
P\Pi_1=\Pi_2,
\end{equation*}
Let $X=diag(\Pi_1,\ I,\ I)$, which is positive definite, so
\begin{equation*}
X^Teqn(11)X < 0,
\end{equation*}
which gives:
\begin{equation}\label{eq:Hinf_LMI_linear} 
\begin{bmatrix}  
    \Pi_1^T(A^TP+PA)\Pi_1&\Pi_1^TPB&\Pi_1^TC^T \\
    B^TP\Pi_1 & -\gamma I & D^T \\
    C\Pi_1 &D&-\gamma I  \\
\end{bmatrix}
  < 0.
\end{equation}
Considering the term $\Pi_1^TPA\Pi_1=\Pi_2^TA\Pi_1,$
\begin{equation}\label{term1}
\begin{split}
\Pi_2^TA\Pi_1=&\begin{bmatrix}
I&0\\S&V
\end{bmatrix}
\begin{bmatrix}
A_p&B_pC_c\\B_cC_p&A_c
\end{bmatrix}
\begin{bmatrix}
R&I\\U^T&0
\end{bmatrix}\\
=&\begin{bmatrix}
A_pR+B_p\hat{C}&A_p\\
\hat{A}&SA_p+\hat{B}C_p
\end{bmatrix},
\end{split}
\end{equation}
where $\hat{A}=SA_pR+SB_pC_cU^T+VB_cC_pR+VA_cU^T$, $\hat{B}=VB_c$, $\hat{C}=C_cU^T$ are the intermittent variables.

Now, the term $\Pi_1^TPB=\Pi_2^TB$
\begin{equation}\label{term2}
\Pi_2^TB=\begin{bmatrix}
I&0\\S&V
\end{bmatrix}
\begin{bmatrix}
B_w\\0
\end{bmatrix}
=\begin{bmatrix}
B_w\\SB_w
\end{bmatrix}
\end{equation}
Now, the term $C\Pi_1$
\begin{equation}\label{term3}
C\Pi_1=\begin{bmatrix}
C_z&0\\
\end{bmatrix}
\begin{bmatrix}
R&I\\U^T&0
\end{bmatrix}
=\begin{bmatrix}
C_zR&C_z
\end{bmatrix}.
\end{equation}
Now, inserting (\ref{term1}-\ref{term3}) back to (\ref{eq:Hinf_LMI_linear}), we can reduce the problem to be LMI,
\begin{equation}
\begin{bmatrix}
\Phi_1+\Phi_1^T&A_p+\hat{A}^T&B_w&R^TC_z^T\\
*&\Phi_2+\Phi_2^T&SB_w&C_z^T\\
*&*&-\gamma I&D_z^T\\
*&*&*&-\gamma I
\end{bmatrix}
\end{equation}
where $\Phi_1=A_pR+B_p\hat{C}$ and $\Phi_2=SA_p+\hat{B}C_p$.

\section{System parameters for the case studies and designed controllers}
For case-study I and II the system parameters are adopted from IEEE Taskforce report on \textit{Benchmark Systems for Small-Signal Stability Analysis and Control} \cite{Xtreport}. The controllers designed for each cases are presented in (\ref{controller_values}) below. 
%\begin{strip}
%\begin{equation*}\label{controller_values}
%\begin{split}
%&\mbox{The controller for four machine system without considering delay,}\\
%&K(s)=\frac{9.42(s+56.39) (s+0.66) (s^2 + 4.17s + 39.55)}{(s+384) (s+0.064) (s+0.0044) (s^2 + 201.3s + 1581)}\\
%&\mbox{The controller for four machine system synchronized feedback considering delay,}\\
%&K(s)=\frac{10.539 (s^2 + 4s + 8)(s^2 + 3.167s + 60.2)  } { (s+5.674)(s^2 + 3.191s + 25.45) (s^2 + 17.02s + 1057}\\
%&\mbox{The controller for four machine system non-synchronized feedback considering delay,}\\
%&K(s)=\frac{8.051(s+0.5824)(s^2 + 1.951s + 14.65)(s^2 + 2.629s + 57.04)} {(s+9.994) (s^2 + 0.5442s + 0.09534)  (s^2 + 1.78s + 32.21)}\\
%&\mbox{The controller for ten machine system without considering delay,}\\
%&K(s)=\frac{91.5(s+9.752) (s+0.9412) (s+4.096) (s^2 + 1.288s + 49.75)  } { (s+6.47) (s+14.94) (s+0.4937) (s+0.000339) (s^2 + 2.597s + 54.62)}\\
%&\mbox{The controller for ten machine system synchronized feedback considering delay,}\\
%&K(s)=\frac{61.88(s+0.4694) (s+24.04)(s-0.02341) (s^2 + 1.554s + 89.82)}{(s+1.991) (s+97.6) (s+47.04) (s+0.1229) (s^2 + 0.1025s + 0.2878)}\\
%&\mbox{The controller for ten machine system non-synchronized feedback considering delay,}\\
%&K(s)=\frac{106 (s+0.08869) (s^2 + 0.7502s + 2.35) (s^2 + 4.073s + 96.51)} {(s+1.867) (s+0.1387) (s+603.4) (s+4.524e04) (s^2 + 1.804s + 76.08)}\\
%\end{split}
%\end{equation*}  
%\end{strip}
\end{document}